\pdfoutput=1
\documentclass[aps,prb,reprint,superscriptaddress,showpacs,showkeys]{revtex4-1}
\usepackage{graphicx}
\usepackage{graphics}
\usepackage{amsmath,amssymb,amsfonts}
\usepackage{colortbl}
\usepackage{times}
\usepackage{array}
\usepackage{bm} 

\renewcommand*{\k}{\mathbf{k}}

\newcommand*{\Hc}{\textrm{H.c.}}

\newcommand*{\ra}{\rangle}
\newcommand*{\la}{\langle}
\newcommand*{\up}{\uparrow}
\newcommand*{\dn}{\downarrow}

\newcommand*{\eps}{\varepsilon}

\renewcommand{\v}{\mathbf{v}}
\newcommand{\x}{\mathbf{x}}
\renewcommand{\S}{\mathbf{S}}
\newcommand{\Om}{\mathbf{\Omega}}

\newcommand{\btau}{\bm{\tau}}
\newcommand*\bsigma{\bm{\sigma}}
\newcommand{\cepd}{Ce$_3$Bi$_4$Pd$_3$}

\newcommand{\red}[1]{\textcolor{red}{#1}}

\begin{document}
\title{Weyl-Kondo semimetals in nonsymmorphic systems} 
\date{\small{\today}}
\author{Sarah E. Grefe}
\email{seg5@rice.edu}
\author{Hsin-Hua Lai}
\affiliation{Department of Physics and Astronomy and Rice Center for Quantum Materials, Rice University, Houston, Texas 77005, USA}
\author{Silke Paschen}
\affiliation{Institute of Solid State Physics, Vienna University of Technology, 1040 Vienna, Austria}
%
\author{Qimiao Si}
\affiliation{Department of Physics and Astronomy and Rice Center for Quantum Materials, Rice University, Houston, Texas 77005, USA}
%
\begin{abstract}
There is considerable current interest to explore electronic topology in strongly correlated metals, with heavy fermion systems providing a promising setting.
Recently, a Weyl-Kondo semimetal phase has been concurrently discovered in theoretical and experimental studies.
The theoretical work was carried out in a Kondo lattice model that is time-reversal invariant but inversion-symmetry breaking.
In this paper, we show in some detail how nonsymmorphic space-group symmetry and strong correlations cooperate to form Weyl nodal excitations with highly reduced velocity and pin the resulting Weyl nodes to the Fermi energy.
A tilted variation of the Weyl-Kondo solution is further analyzed here, following the recent consideration of such effect in the context of understanding a large spontaneous Hall effect in Ce$_3$Bi$_4$Pd$_3$ 
(Dzsaber \textit{et al.}, arXiv:1811.02819).
We discuss the implications of our results for the enrichment of the global phase diagram of heavy fermion metals, and for the space-group symmetry enforcement of topological semimetals in other strongly correlated settings.
\end{abstract}
\maketitle
%
\section{\label{sec:intro} Introduction}

\begin{figure}[t]
   \centering
		\includegraphics[width=.49\columnwidth]{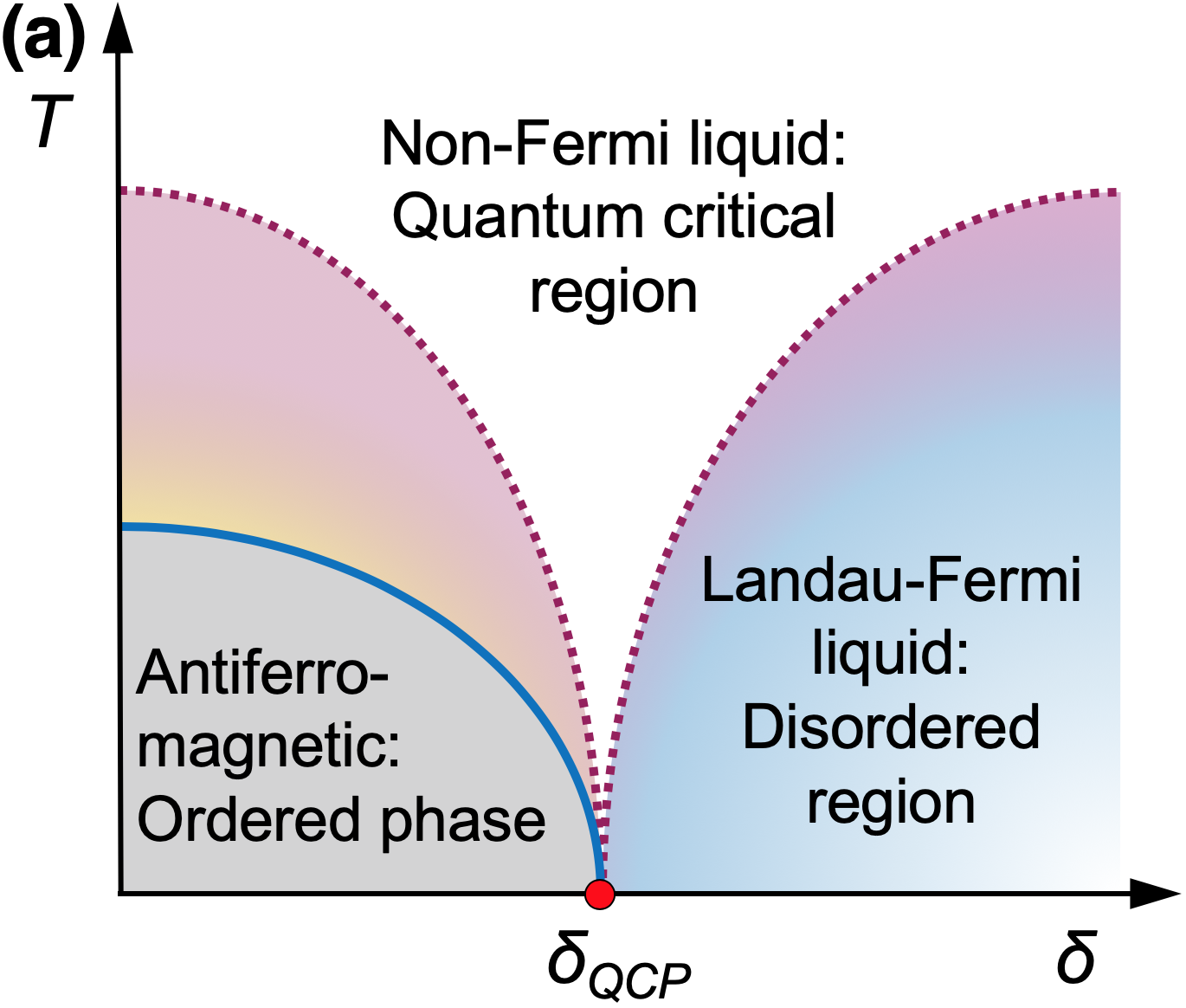}
		\includegraphics[width=.49\columnwidth]{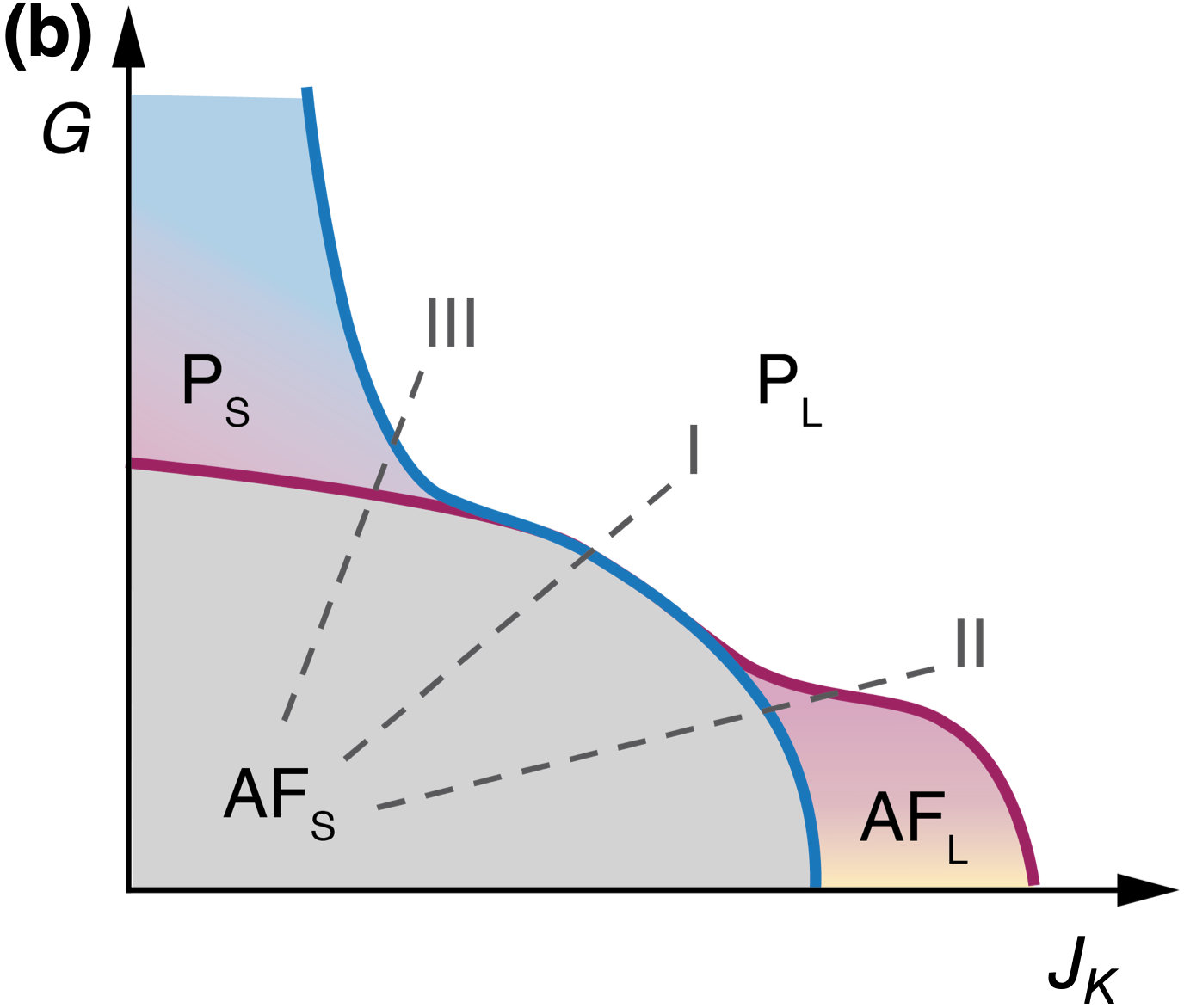}
   \caption{ \label{fig:qpt-hf} 
   (a) Quantum phase transition of heavy fermion metals.
      The vertical axis varies temperature $T$ and thus the amount of thermal fluctuations.
      The horizontal axis tunes a non-thermal control parameter $\delta$, which corresponds to the ratio of the Kondo to RKKY interactions. The red point labeled $\delta_{\text{QCP}}$ marks a quantum critical point where the ordered antiferromagnetic phase and the disordered heavy Landau-Fermi liquid phases meet at $T=0$, producing the quantum critical non-Fermi liquid regime at nonzero temperatures.
     (b) Heavy fermion global phase diagram.~\protect\cite{SI200623,Si_PSSB10,Custers2012,Jia15,Luo:2018aa,Zha19}
   The parameter $G$ controls the amount of geometric frustration of local moments, while $J_K$ modifies the magnitude of the Kondo coupling. 
	AF labels phases with antiferromagnetic ordering of local moments, and P labels paramagnetic phases. Subscript $L$ denotes large Fermi surface phases, and subscript $S$ denotes small Fermi surface phases. 
	The dashed lines, labelled ``I'', ``II'' and ``III'',
	 illustrate three trajectories (corresponding to three cuts in the parameter space for the tuning 
	parameter $\delta$ shown in Fig.~\ref{fig:qpt-hf}(a)) of quantum phase transitions.
   Near the border of the AF phase boundaries, large magnetic fluctuations may lead to emergent correlated topological phases in heavy fermion systems when the SOC is large.
      }
\end{figure}

Strong correlations give rise to a plethora of ground states and, correspondingly, a variety of quantum phase transitions. 
Heavy fermion metals have provided a canonical setting to study strong correlation physics, 
including quantum critical points and emergent  phases.~\cite{si_QPT_rev10,HvL2007,Stewart2001} 
Typically, in these systems, $4f$ electrons have a local Coulomb repulsion that is large compared to their bandwidth.~\cite{hewson_book} 
Due to such strong correlations, the $4f$ electrons act as local moments, which are Kondo-coupled to a band of background conduction electrons. 
The local moments can form a Kondo singlet with the spins of the conduction electrons by the Kondo coupling, or they may condense into an antiferromagnetic order through their Ruderman-Kittel-Kasuya-Yosida (RKKY) interaction.~\cite{doniach77} 
Correspondingly, antiferromagnetic quantum critical points often develop in heavy fermion metals, as illustrated in  Fig.~\ref{fig:qpt-hf}(a).
A global phase diagram has been advanced~\cite{SI200623,Si_PSSB10,Custers2012,Jia15,Luo:2018aa,Zha19} 
as shown in Fig.~\ref{fig:qpt-hf}(b). 
It features quantum phases that are distinct not only by the existence or absence of antiferromagnetic order, but also by the Kondo entanglement and its destruction.~\cite{Si-Nature, Colemanetal, senthil2004a, Schroder, paschen2004, Gegenwart2007, Friedemann.10, Martelli17701, shishido2005}

An intriguing problem is how the overall quantum phase diagram of heavy fermion metals is enriched by phases that are topologically non-trivial, when the strong correlations interplay with a large spin-orbit coupling.~\cite{SiPaschen2013} 
This is an outstanding issue at the intersection between strong correlations and topology.
From the vantage point of non-interacting topological states of matter, the typical questions of interest
concern the effect of electron correlations on non-interacting topological states, such as non-interacting topological semimetals, 
or about the type of weakly interacting topological phases that can be produced by interaction-induced broken symmetries.~\cite{Gonzalez-Cuadra:2019aa,PhysRevB.92.075438,PhysRevLett.109.066401}
However, in the present context, the question of interest is different: What type of topological metallic states can be driven by strong correlations? 
We address this issue in a setting where the time-reversal symmetry is preserved.

Recently, a Weyl-Kondo semimetal (WKSM) phase has been concurrently discovered in theoretical~\cite{WKSM_PNAS} and experimental studies.~\cite{DzsaberPRL2017,2018arXiv181102819D}
This theoretical work was carried out in a Kondo lattice model that is time-reversal invariant but inversion-symmetry breaking. 
The defining characteristics of the Weyl-Kondo semimetal include linearly-dispersing Weyl nodal excitations with highly reduced velocity and Weyl nodes being pinned to the Fermi energy.

In this paper, we show in some detail the role of nonsymmorphic space group symmetry in producing these properties.
We consider a periodic Anderson/Kondo model on a diamond lattice, with inversion symmetry broken by a staggered potential, at quarter filling.~\cite{WKSM_PNAS}
Focusing on the limit of large on-site Coulomb repulsion, the model is equivalent to a Kondo lattice. 
In the absence of Kondo coupling, the nonsymmorphic space group symmetry generates Weyl nodes that are located far away from the Fermi energy, and the ground state is topologically trivial. 
Because of the Kondo effect, strongly renormalized quasiparticles are produced near the Fermi energy. 
When this happens, the space-group symmetry in turn ensures that the Weyl nodes develop precisely at the Fermi energy; 
this makes the Weyl nodal excitations to be long-lived and, hence, well-defined, even in the present strongly interacting setting.
In addition, the renormalized nodal velocity is smaller than the usual non-interacting value by the ratio of the Kondo temperature to the bare conduction-electron bandwidth, which can be as large as three orders of magnitude.
We will also analyze further the effect of a tilting potential~\cite{2018arXiv181102819D} to the Weyl-Kondo solution itself and the Berry curvature distribution. All these properties are important in giving rise to new signatures of the Weyl-Kondo semimetal in thermodynamic~\cite{WKSM_PNAS,DzsaberPRL2017} and
transport properties.~\cite{2018arXiv181102819D} 
In addition, we will discuss how these results enrich the global phase diagram of heavy fermion metals. 
This enrichment captures the role of spin-orbit coupling in the interplay between competing phases, all of which develop out of the underlying spin degrees of freedom of the $4f$ electrons.
Finally, we touch upon the implications of the results for space-group symmetry enforcement of topological semimetals in the general context of  strongly correlated systems.

To set the stage for our analysis about how the space-group symmetry interplays with strong correlations, we start by briefly outlining the role of space-group symmetry in the noninteracting case.

\subsection{\label{sec:crystal} Role of nonsymmorphic space group symmetry and protection of topological semimetal phases } 

We focus our discussion on three dimensional (3D) crystals.
In topological semimetals, the bulk already has a gapless excitation spectrum. 
This is to be contrasted with topological insulators, in which the bulk excitations are fully gapped and only the surface states are gapless.
Both of these topological phases must have band inversion, a reordering of conduction and valence bands which allows the topological insulator surface states to connect the conduction and valence bands.
In three dimensions, a quadratic Hamiltonian can be classified into a topological equivalence class depending on its nonspatial symmetries: time-reversal symmetry (TRS), particle-hole symmetry, and chiral symmetry.
This is commonly known as the tenfold way, which can be applied to topological insulators and topological semimetals, and can accommodate spatial symmetries as well.~\cite{RevModPhys.88.035005}
More specifically to Weyl semimetals, these are characterized by energy levels that meet in pairs of twofold degenerate points in momentum space as a result of tuning Hamiltonian parameters.
Without other symmetries in the system, this requires tuning three parameters to achieve degeneracy, but space group symmetries may protect the degeneracies.~\cite{AMV_RMP}
In the efforts to identify topological materials, several studies have applied symmetry classification criteria to space groups~\cite{SlagerPRX2017,Murakami2017,po2017symmetry} and more specifically to particular lattice realizations,~\cite{bradlyn2017topological,Cano2018} 
as well as considered the fillings at which nonsymmorphic symmetries will enforce gapless phases.~\cite{Watanabe2016}

To obtain such degeneracy, a mechanism of band inversion is a necessary but insufficient ingredient.
Band inversion can occur by lattice strain, scalar relativistic effects, or spin-orbit coupling (SOC).~\cite{PhysRevB.85.235401,PhysRevLett.95.146802,PhysRevLett.95.226801}
Given the crucial role of the SOC, the search for topological materials tends to focus on systems that are based on heavy elements with large SOC.
Conveniently, the lanthanides and actinides where the $f$-orbital elements as well as the often-involved heavy elements (e.g., Bi) associated with heavy fermion materials provides substantial SOC.

We are interested in topological semimetals in 3D crystals with SOC and additional space-group symmetries that can protect nodal band crossings.
It turns out that many nonsymmorphic space groups can support four-dimensional irreducible representations on the zone boundaries, which produce robust symmetry protected Dirac semimetal phases, provided they do not lie along threefold or sixfold rotation axes.~\cite{AMV_RMP,Kane_3ddirac}
A nonsymmorphic symmetry is a space group operation $\{\mathcal{O}|\mathbf{t}\}$ which combines a spatial point-group operation (or nonspatial operation) $\mathcal{O}$ with a \textit{partial} (non-primitive) lattice translation vector $\mathbf{t}$.
	Spatial symmetries are group operations that rotate and reflect different lattice sites onto one another, such as an $n$-fold rotation about the $i$th axis $C_{ni}$, or a reflection about the $ab$ plane with normal vector $\hat{c}$, $m_{c}$.	
	Respectively, the corresponding nonsymmorphic transformations are called screw operations ($\{C_{ni}|\mathbf{t}\}$ = rotation + fractional $\mathbf{t}$) and glide operations ($\{m_c|\mathbf{t}\}$ = reflection + fractional $\mathbf{t}$).
Since multiple fractional translations are needed to traverse the unit cell, it is enlarged in real space, which causes the Brillouin zone (BZ) to fold.
This creates a new Brillouin zone boundary (BZB) where any bands that intersect it are sharply reflected back into the BZ, causing a degeneracy at the BZB.
For these reasons, nonsymmorphic space groups with SOC generically produce Dirac nodal band touching points or lines of degeneracy. 
The glide symmetry is familiar to the strongly correlated electron community of iron pnictides; there, the symmetry implies that the eigenstates come in (glide even and odd) pairs and, as a result, gives rise to an extra degeneracy at the boundary of the BZ associated with the physical two-iron unit cell.~\cite{Nica2015}

With the nonsymmorphic symmetry-enforced Dirac semimetal as a starting point, a Weyl semimetal phase can arise from breaking TRS or inversion symmetry (IS).~\cite{AMV_RMP}
Without the protection of space-group symmetry, one would have to resort to IS breaking (ISB) systems tuned to within a band inversion transition between a trivial band insulator and a topological insulator.~\cite{Murakami2017,PhysRevB.78.165313}
Bands that invert are allowed to cross because 
(1) the bands have different irreducible representations, such as the odd-even parity in $s$-$f$ coupling; 
(2) bands of the same irreducible representation may have wavefunctions that differ by a Berry phase.~\cite{RevModPhys.88.035005,SchnyderTopoLecture}
Otherwise, the noncrossing theorem requires that the bands hybridize to open a topologically trivial gap at a generic point in the BZ.
Thus a robust procedure is to search for space groups that anchor IS breaking or are also noncentrosymmetric, and can realize an even filling factor that is both gapless and has zero enclosed Fermi surface volume.
	
\subsection{\label{sec:top_strong_corr} Topological states driven by strong correlations}

Given all these considerations, a topological state driven by the Kondo effect arises if one first realizes a topologically 
trivial ground state without the Kondo effect, and when the Kondo effect is turned on, produces a topologically non-trivial phase.
Our model has a solution that corresponds to such a Kondo-driven phase. 
We will show that the result is robust to changes in parameters. This is because our Hamiltonian has the required crystal and local symmetries, and fulfills group-theoretical filling constraints 
that achieve topological semimetal phases in response to the Kondo effect.
Therefore, our model illustrates that strong correlations help hone in on nontrivial topological phases in the vast multidimensional parameter space of the strong correlation global phase diagram
in the presence of a large SOC.
In this sense, a design principle follows from our work (as well as from experiments~\cite{DzsaberPRL2017,2018arXiv181102819D}),
namely to search for topological semimetals driven by strong correlations by focusing on strongly correlated semimetals with a nonsymmorphic space group and broken inversion symmetry.

The remainder of the paper is organized as follows. 
In Sec.~\ref{sec:model}, we first explain our model and solution method.
In Sec.~\ref{sec:wksm}, we show how the model obtains a correlations-driven topological phase transition 
from specific symmetry considerations (Sec.~\ref{sec:realization}) and the pinning of the nodes to the Fermi energy driven by the combined effect of the space-group symmetry and strong correlations (Sec.~\ref{sec:knode}). 
A symmetry-allowed tilt term is added in the model and its role is analyzed in some detail in Sec.~\ref{sec:tilt}.
Then in Sec.~\ref{sec:sigs}, we discuss strong correlations-suited transport and thermodynamic signatures of WKSM phases.
Finally, we close in Sec.~\ref{sec:conclusion} by discussing heavy fermion materials as a platform for exploring topological phases,  and consider the nature of topological phase transitions in the global phase diagram of heavy fermion metals.

\section{\label{sec:model} Model and solution method}

To achieve the WKSM phase, we used the following model~\cite{WKSM_PNAS} defined on the diamond lattice:
\begin{align}
\mathcal{H} &= \mathcal{H}_c + \mathcal{H}_{cd} + \mathcal{H}_{d}.\label{eq:hamiltonian}
\end{align}
This is an Anderson lattice type model which describes the heavy fermion systems.
The Hamiltonian separated into the part representing physical conduction $spd$ electrons $\mathcal{H}_c$, a hybridization term $\mathcal{H}_{cd}$ which allows the formation of heavy yet mobile quasiparticles, and a part representing the physical highly localized $4f$ electrons $\mathcal{H}_d$.

The conduction electrons are described by
\begin{align}
	\mathcal{H}_c &= t\sum_{\la i j \ra, \sigma } \left( c_{i \sigma}^\dagger c_{j \sigma}+ \Hc \right)\nonumber\\
	&- \mu \sum_{i, \sigma} n_{i \sigma}^c\nonumber\\
	&+ i \lambda \sum_{\la\la i j \ra\ra} \left[ c^\dagger_{i \sigma} \left( \bsigma\cdot {\bf e}_{ij} \right) c_{j\sigma} - \Hc\right]\nonumber\\
	&+ m \sum_{i,\sigma} (-1)^i c^\dagger_{i\sigma} c_{i\sigma}.
	\label{eq:condel}
\end{align}
This Hamiltonian is based on the Fu-Kane-Mele 
model.~\cite{FKMmodel07,Ojanen13,PhysRevB.78.165313}
The nearest-neighbor ($\la i j \ra$) hopping amplitude sets the energy scale at $t=1$; the chemical potential parametrizes the electron density as $\mu$ and breaks particle-hole symmetry; the second-nearest neighbor ($\la\la i j \ra\ra$) Dresselhaus-type spin orbit coupling has strength $\lambda$ 
and acts upon spin space in the Pauli matrix basis, in vector form $\bsigma$;
finally a sublattice-dependent atomic potential 
difference is tuned by $m$. 
In this last term, $m>0$ tunes the degree of crystal inversion symmetry breaking, serving to lift the degeneracies between the two face-centered-cubic (fcc) sublattices of the diamond lattice.

The term that allows the two species to hybridize is simply
\begin{align}
\mathcal{H}_{cd} &=
V \sum_{i,\sigma} \left( d_{i \sigma}^\dagger c_{i \sigma} + \Hc\right),\label{eq:hybridization}
\end{align}
which has strength $V$, and in the strong coupling limit
that we consider, tracks the strength of the Kondo effect.

To represent the localized $4f$ electrons, the $d$-operator Hamiltonian is
\begin{align}	
	\mathcal{H}_d&=E_{d} \sum_{i, \sigma} d_{i \sigma}^\dagger d_{i \sigma} + U \sum_i n^d_{i \up} n^d_{i \dn},\label{eq:hd}
\end{align}
The first term has a flat bare atomic energy level $E_d$ which lies far below the conduction electron bands.
The second term is the Coulomb interaction with repulsion energy $U$, which penalizes the double occupation of a site.

Finally, we will primarily focus on the case of quarter filling, which corresponds to total electron count of $1$ per site:
\begin{align}
\label{eq:filling}	
	n_d + n_c = 1,
\end{align}
where, 
\begin{align}
	n_d&=\frac{1}{N_\text{site}}\sum_{i,\sigma } d_{i \sigma}^\dagger d_{i \sigma},\\
	n_c&=\frac{1}{N_\text{site}}\sum_{i,\sigma } c_{i \sigma}^\dagger c_{i \sigma},
\end{align}
with $N_\text{site}$ counting the total number of sites in
the lattice.

The interaction $U$ term is, of course, an obstruction to obtaining the eigenstates.
However, in the strong coupling limit of $U\rightarrow\infty$, one can use the auxiliary boson method~\cite{hewson_book}
to treat the Coulomb term by considering it's large limit consequence, which is to only allow density configurations of single particle-per-site occupation and empty occupancy.
Thus the localized species acquires a boson as $d^\dagger_{i\sigma}=b_i f^\dagger_{i\sigma}$, which is, at the saddle-point level,
 averaged over the unit cell as $b_i\rightarrow\la b_i \ra=r$, where $0<r<1$.
This necessitates including a Lagrange multiplier $\ell$ which parameterizes a constraint equation term introduced into the large $U$ Hamiltonian as $\mathcal{H}_s=\mathcal{H}+\mathcal{H}_\ell$,
\begin{equation}
	\mathcal{H}_\ell=\ell\left(\sum_{i,\sigma}f^\dagger_{i\sigma}f_{i\sigma}+r^2-1\right),
\end{equation}
and renormalizing the hybridization as $V\rightarrow\tilde{V}=rV$.
Put together, this gives the strong coupling Hamiltonian $\mathcal{H}_s$.

The parameters $\x=(\mu,r,\ell)$ are obtained by solving the set of saddle point equations $\frac{\delta \mathcal{H}_s}{\delta x_i}=0$ self consistently.
The parameter $\ell$ renormalizes the localized electron's energy level to $E_d\rightarrow \tilde{E}_d=E_d+\ell$, which in practice is close to $E_F$ (we define $E_F=0$).

We eliminated the need to numerically solve for $\mu$ by finding the analytical solutions to nodal points in the Brillouin zone (see Appendix~\ref{app:pmD}).
The key step to solving for the eigenenergies is to find a suitable basis that renders the Hamiltonian separable.
In Ref.~\onlinecite{WKSM_PNAS}, we performed the canonical (unitary) transformation on Eq.~(\ref{eq:condel}), $\breve{\Psi}_\k = S^\dagger_\sigma\Psi_\k$, that leads to
\begin{align}
 H_c &= \sum_\k \breve{\Psi}^\dagger_\k \begin{pmatrix} 
 	h_{\k +} & 0 \\ 0 & h_{\k-} \end{pmatrix} 
	\breve{\Psi}_\k,\\
h_{\k \pm} &= u_1(\k)\tau_x + u_2(\k)\tau_y + ( m \pm \lambda D(\k))\tau_z,\label{eq:hminus}
\end{align}
where $u_1(\k),~u_2(\k),~D(\k)$ are defined for the diamond lattice in Appendix~\ref{app:pmD}, and the $\tau_i$ are Pauli matrices acting in the fcc sublattice space.
	We have used a pseudospin basis,~\cite{Ojanen13} defined by the eigenstates $|\pm\,D\ra$ with eigenvalues 
\begin{align}
	&\frac{\bm{D}(\k)\cdot \bsigma}{D(\k)}|\pm D\ra=\pm|\pm D\ra,\label{eq:pseudospin}\\
	&D(\k) \equiv \left| \bm{D}(\k) \right| = \sqrt{ D_x(\k)^2 + D_y(\k)^2 + D_z(\k)^2}\label{eq:Dk}
\end{align}
Based on previous studies,~\cite{Ojanen13,WKSM_PNAS} we know that the Weyl nodes only emerge in the $|-D\ra$ sector corresponding to $h_{\k -}$. 
The term $D(\k)$ arises from the Fourier transform of the SOC term.
Since the bands have a definite ordering in terms of energy (see Appendix~\ref{app:pmD}), we find the nodal band touchings occur only between particular bands (see Sec.~\ref{sec:knode}).

With the number of particles per-site-per-spin (or simply fractional filling) of all fermions being $n_c+n_d=1$ and since the localized electrons' filling was fixed at $n_f=1-r^2$, this implies that the conduction electron density is $n_c=r^2$ and thus small.

\begin{figure}[h]
   \centering
   		\includegraphics[width=.75\columnwidth]{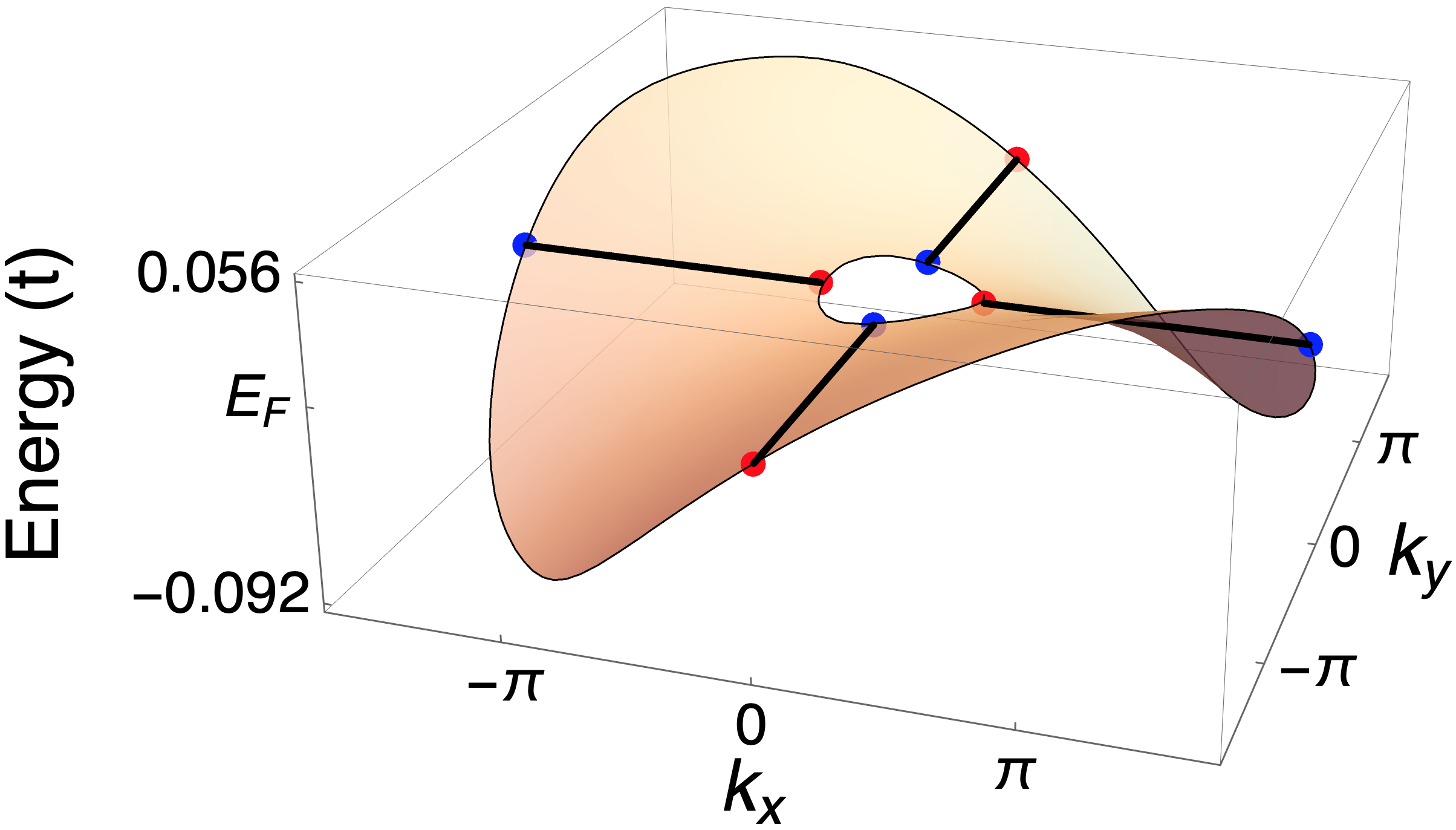}
	\caption{ \label{fig:surfacestates} Eigenenergy of the surface states of the $[001]$ plane, with parameters $(E_d,\ell,r,V)=(-7,7.28,0.22,7.5)$.
	Blue and red points show the position of the Weyl and anti-Weyl nodes of the Brillouin zone boundary, and thick black lines show the Fermi arcs connecting nodes to their opposite chirality partner on the four neighboring Brillouin zone boundaries.}
\end{figure}

\section{\label{sec:wksm} Weyl-Kondo semimetal}

In Ref.~\onlinecite{WKSM_PNAS} we established that our model captures a Weyl Kondo semimetal featuring Kondo renormalization-narrowed bulk and surface bandwidths, exhibiting bulk Weyl nodes and surface states with Fermi arcs.
An example of the surface states and their band narrowing can be seen in Fig.~\ref{fig:surfacestates}, in which the bandwidth of the pure surface states 
is renormalized by the Kondo effect.

The configuration of Weyl monopoles is tuned along the Brillouin zone \textit{boundaries} with a quarter of the bands filled (two of eight total), which is consistent with our choice of nonsymmorphic lattice and Hamiltonian.~\cite{1367-2630-9-9-356,Murakami2017,Kane_3ddirac}
The Weyl-Kondo quasiparticles form exactly at the Fermi energy, for reasons at two levels of sophistication 
\begin{itemize}
\item[(i)]
 the Weyl nodes appear within the Kondo resonances, which lie near the Fermi energy within a small energy window set by the Kondo temperature and 
\item[(ii)]
 the space group symmetry combined with the commensurate filling puts the nodes even closer to the Fermi energy - for the exact commensurate filling, the nodes are precisely at the Fermi energy.
 \end{itemize}

In other words, the combination of the Kondo effect and space group symmetry pin the Weyl nodes to the Fermi energy.

Here, we analyze the mechanism that underlies this salientfeature of the Weyl-Kondo semimetal phase.

\begin{figure*}[t]
   \centering
      	\includegraphics[width=.51\columnwidth]{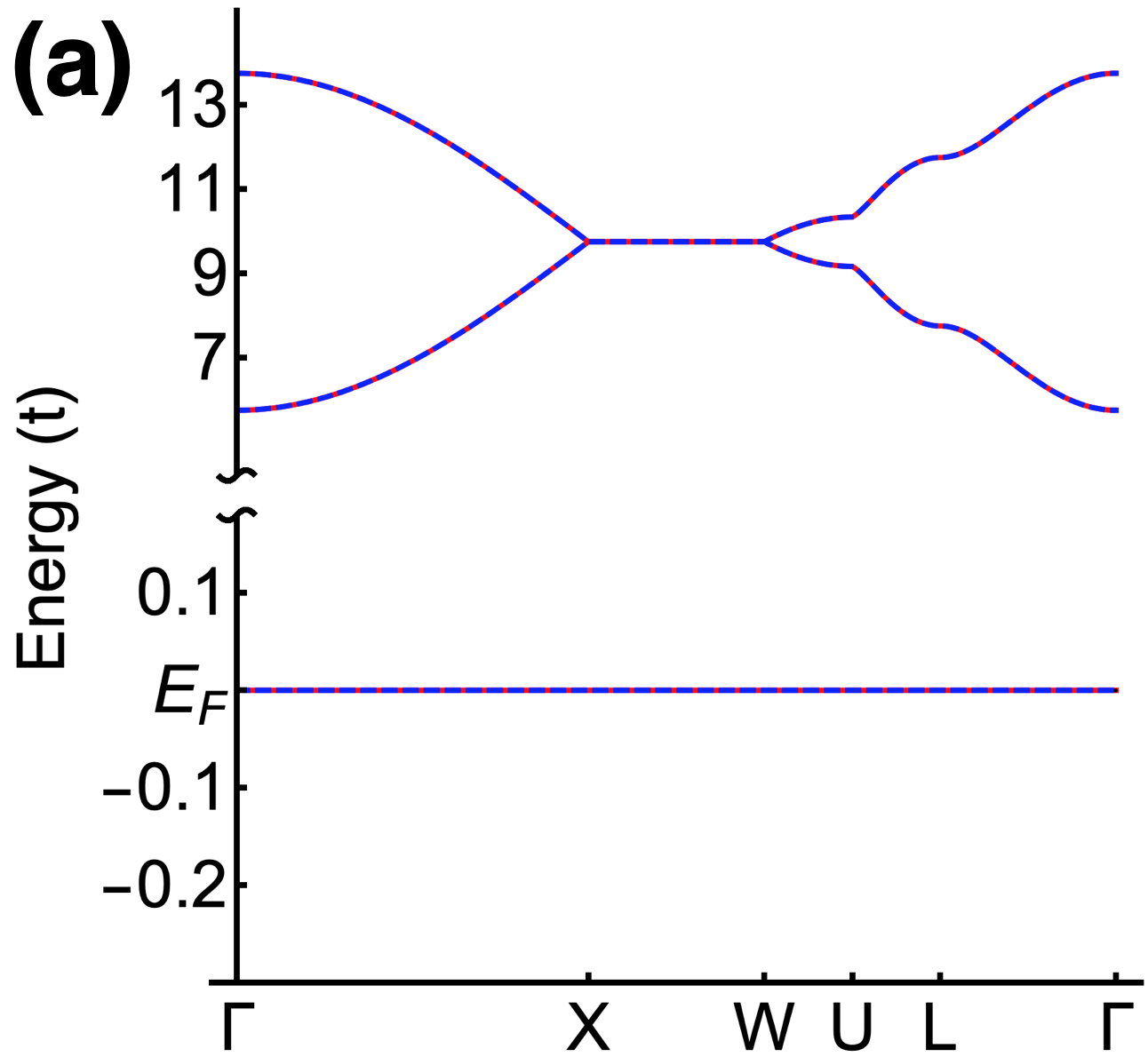}~~~~
   		\includegraphics[width=.51\columnwidth]{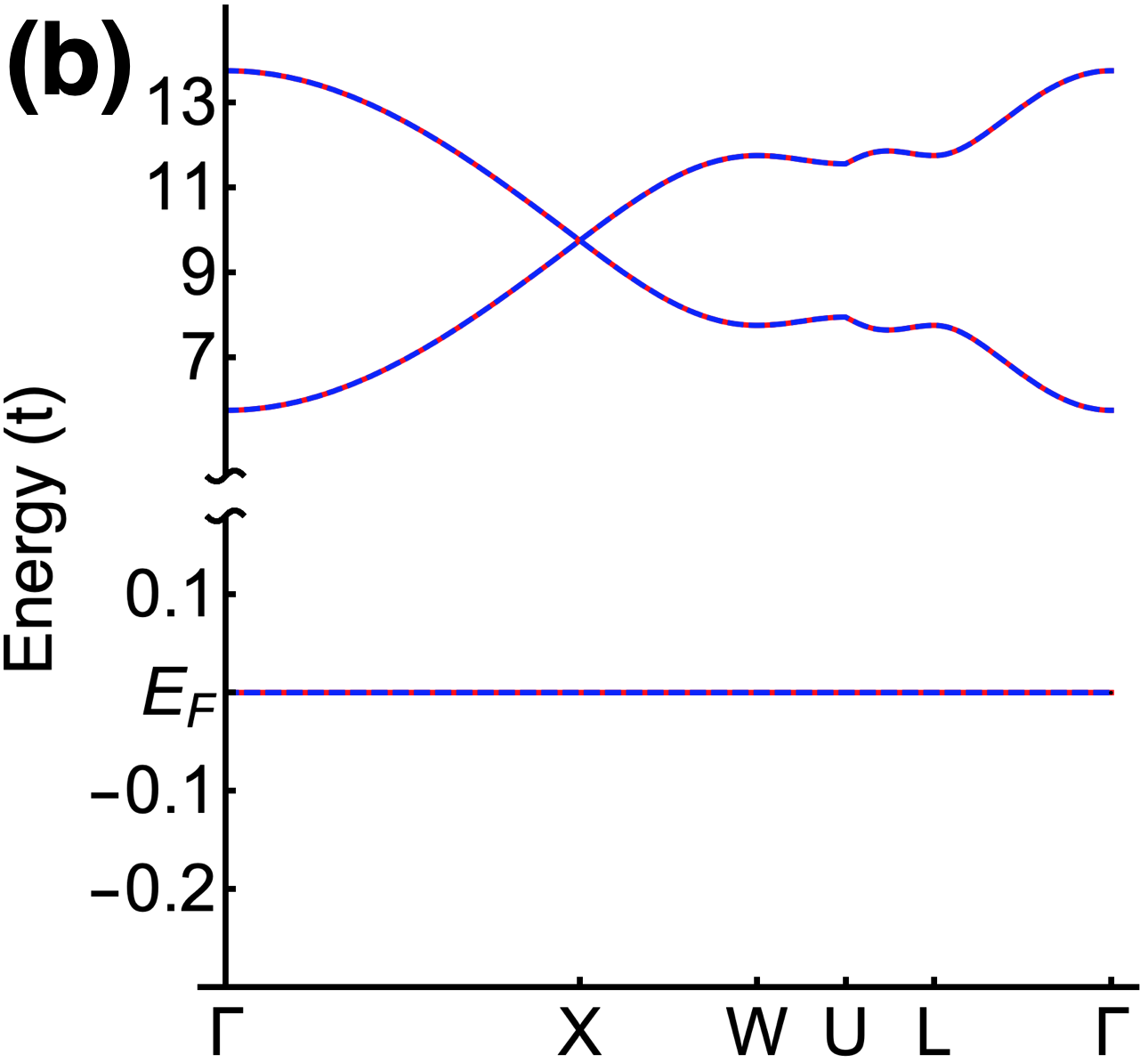}~~~~
		\includegraphics[width=.51\columnwidth]{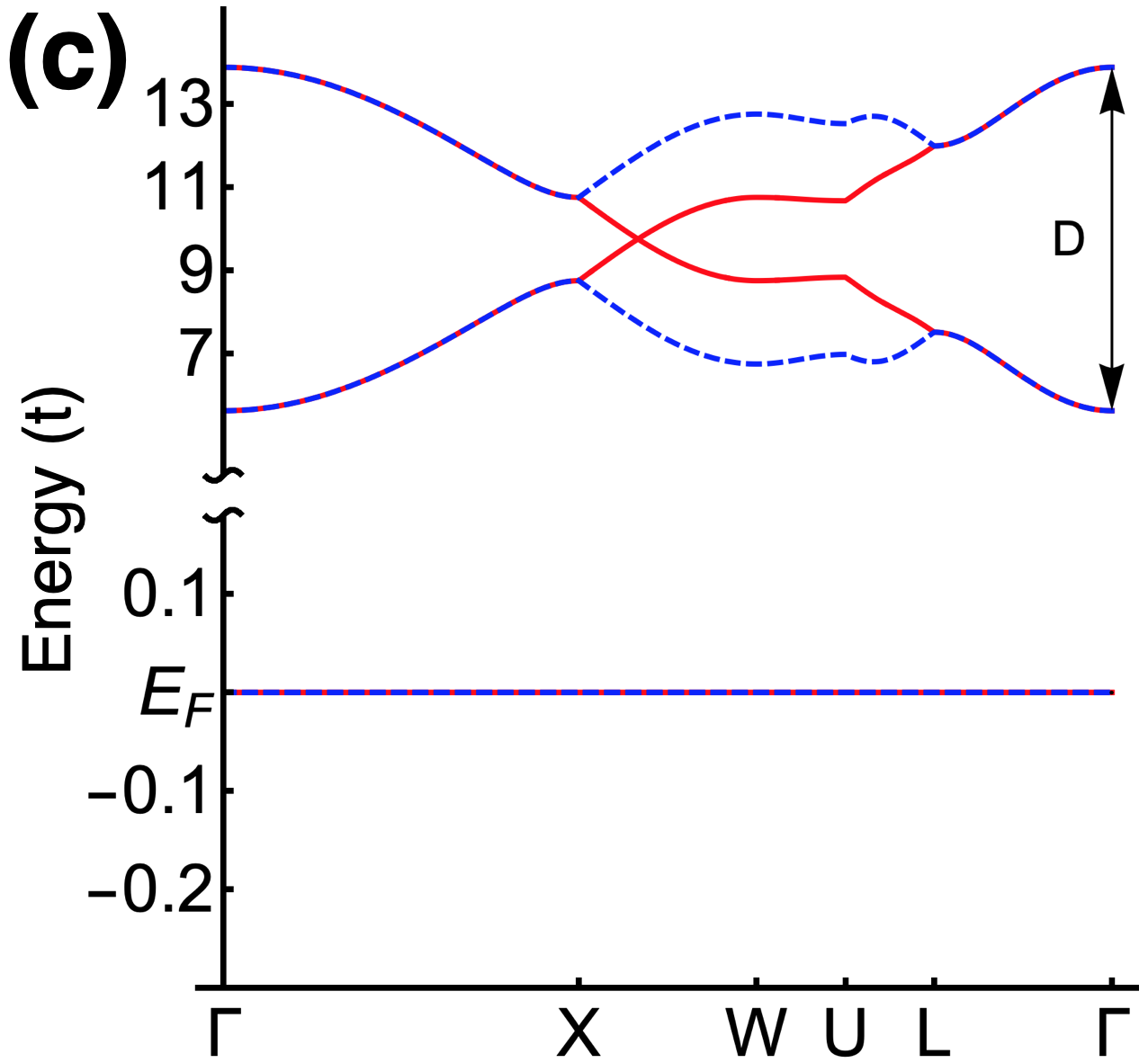} 
   \caption{ \label{fig:still-trivial}
   \textit{Model in the absence of the Kondo effect, $V=0$.} Dispersion along a high symmetry path in the fcc Brillouin zone.
   The localized electrons' energy level is at the Fermi energy $E_F$.
   (a) Fourfold-degenerate line nodes along $X-W$ when SOC $\lambda=0$ and inversion symmetry is preserved $m=0$.
   (b) Dirac node develops at $X$ from (a) when SOC is present, $\lambda>0$. 
   (c) Weyl nodes emerge from (b) when inversion symmetry is broken and $0<\frac{m}{4\lambda}<1$.
   $D\sim8t$ denotes the unrenormalized conduction electron bandwidth.
	} 
\end{figure*}
\subsection{\label{sec:realization} Realization of the Weyl-Kondo semimetal through symmetry}

In Sec.~\ref{sec:crystal}, we presented some considerations for seeking Weyl semimetals in 3D crystal lattices.
We now turn to the specific case of the diamond lattice.

The diamond lattice has space group no. 227 ($Fd\bar{3}m$), which is centrosymmetric, nonsymmorphic, and bipartite, consisting of two fcc lattices displaced by ${\mathbf t}=a\{\tfrac{1}{4}\tfrac{1}{4}\tfrac{1}{4}\}$ ($a$ is the cubic lattice constant).
	First, consider the Hamiltonian \textit{without Kondo coupling  ($V=0$)} implemented on the diamond lattice, which involve four bands associated with two sublattices and two spin states, with localized $E_d$ levels completely decoupled from them.
	We consider the symmetry change as particular terms are successively added.
	In Fig.~\ref{fig:still-trivial}, the cases without Kondo coupling are shown, where the flat, trivial $f$ bands are well separated in energy below the dispersive $c$ bands.
	For the quarter-filling case we consider, the localized $d$ electrons are at half filling, forming a Mott insulator, and the conduction $c$-electron bands are left completely empty, implying a topologically trivial insulator phase.
	If only the nearest-neighbor hopping and chemical potential terms [first two lines of Eq.~(\ref{eq:condel})] are included, one has line nodes along $X-W$ as shown in the upper set of bands in Fig.~\ref{fig:still-trivial}(a).
	These fourfold degenerate line nodes crisscross the square BZB, due to a combination of the nonsymmorphic symmetry,
	two of the mirror planes, and the $C_4$ and $C_2$ rotations.

Next, we include the SOC (which preserves TRS and IS), which is shown in Fig.~\ref{fig:still-trivial}(b).
	The fourfold degeneracy at $W$ is split, while a linearly dispersing degenerate Dirac point remains at $X$.
	The Dresselhaus SOC term which allows the pseudospin $|\pm D\ra$ decomposition also allows the band inversion by introducing a linear-in-$\k$ coupling term based on the pseudospin eigenvalue $D(\k)$, which is linear near the $X$ point,
\begin{align}
	D\left[\k_X=(k_0,0,2\pi)\right]&= 4\sqrt{\sin^2\left(\frac{k_0}{2}\right)} \sim 2|k_0|
\end{align} 
for some $k_0\sim0$, and similarly for the other $X$ points.
The bands are twofold degenerate everywhere in the BZ except at the Dirac points, and in Fig.~\ref{fig:still-trivial} this is indicated with solid red lines for the $|-D\ra$ sector, and dashed blue lines for the $|+D\ra$ sector.
Since the SOC preserves TRS, the Kramers degeneracy at the time-reversal invariant momentum $X$ is preserved, while the SOC splits them at $W$ since it is not a time-reversal invariant momentum point.
The space group analysis of how the nonsymmorphic symmetry of the diamond lattice produces such Dirac nodes at the $X$ points have been established previously in a noninteracting model without a localized species; 
on the BZB, a projective representation with point group $D_{4h}$ has a four-dimensional irreducible representation, which realizes a Dirac semimetal generically.~\cite{Kane_3ddirac}

We now include the ISB term parametrized by $m$, shown in Fig.~\ref{fig:still-trivial}(c). 
The $|\pm D\ra$ degeneracy is split along the BZB, and doubly degenerate Weyl nodes of the $|-D\ra$ sector emerge along all $X-W$ lines.
This degeneracy produced by the internal sublattice degree of freedom is lifted, but since TRS is preserved, the Kramer's pairs remain with their TRS partners at $X$.
This allows one to tune $\k$ between the $X$ and $W$ points to find a Weyl node degeneracy.

Put a different way, one can track the $X-W$ degeneracies as a function of the IS breaking.
The Weyl semimetal phase region is $0<m<4\lambda$, with the Dirac node at $X$ ($m=0$) splitting and spawning the Weyl nodes as $m$ is increased.
The four nodes move outward toward each of the four $W$, and undergo a quadratic band touching at the critical value $m=4\lambda$, before annihilating with the nodes of opposite chirality from the four neighboring BZs, which opens a trivial gap when $m>4\lambda$.~\cite{PhysRevB.78.165313,Ojanen13}
	
It is also pertinent to consider the space group symmetry when identifying which pairs of bands can form band touching points, that is, to find what filling factor realizes a topological semimetal for a given space group.
The filling factor $\nu$ counts the number of electrons per primitive unit cell; in our model [Eq.~(\ref{eq:hamiltonian})], there are two types of fermions, two spins per fermion, and two sites per unit cell, so the total allowed filling factor is $\nu=8$. 
At $\nu=2$ corresponding to quarter-filling, the nonsymmorphic symmetry enforces that both the Fermi surface must be finite, and yet the Luttinger volume must vanish.~\cite{Watanabe2016}
The only way to satisfy these conditions is to produce a zero dimensional nodal point Fermi surface, 
so nonsymmorphic space groups are a natural place to search for topological systems.
When the space group symmetry is changed via ISB, this restores the ability of the system to connect adiabatically to a band insulator phase, when $m>4\lambda$.

\subsection{\label{sec:knode} Kondo-driven node formation and pinning}

\begin{figure}[t]
		\includegraphics[width=.6\columnwidth]{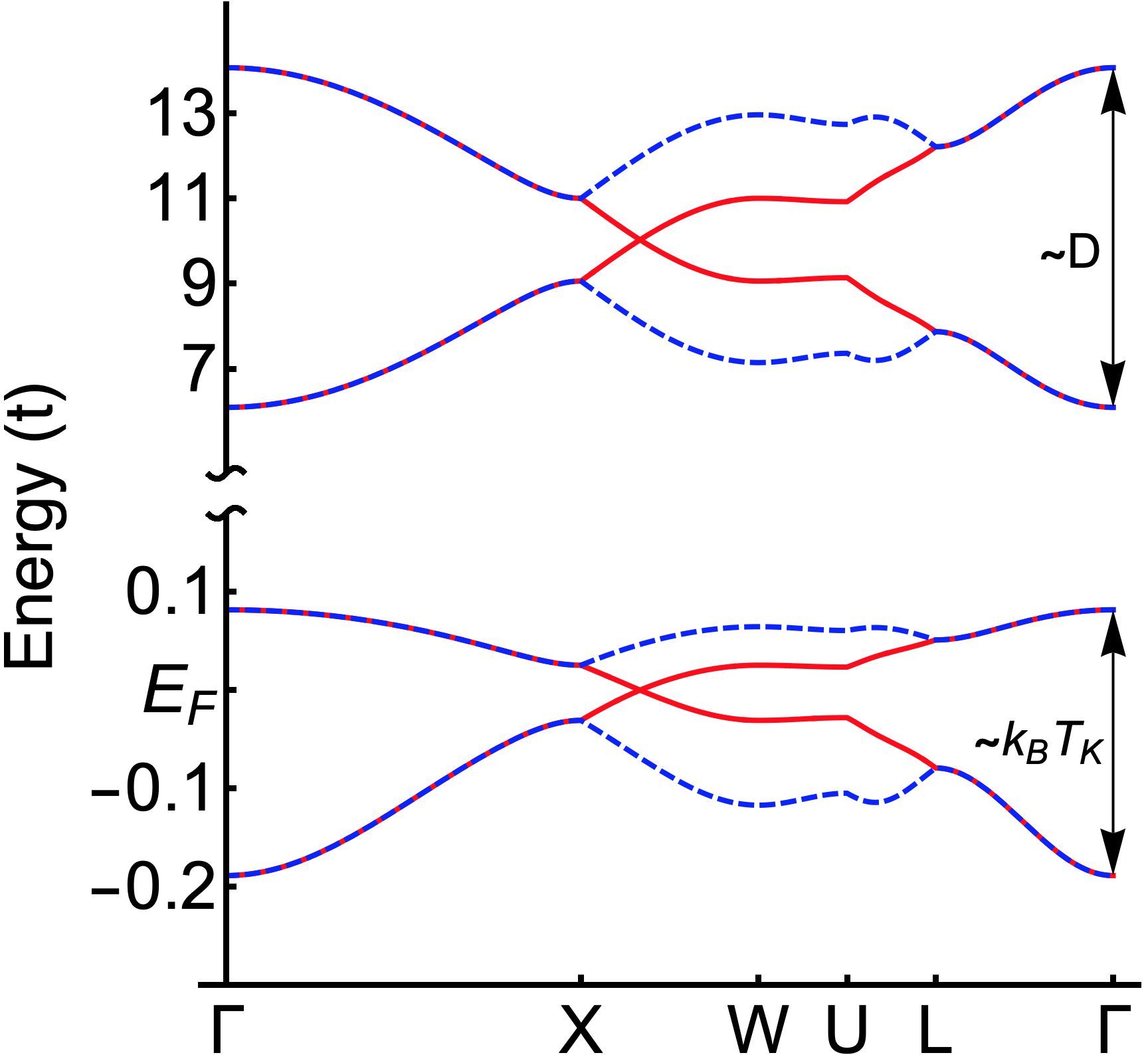}
   \caption{ \label{fig:kondopin}
\textit{Kondo effect-driven Weyl nodes.} Dispersion along a high symmetry path in the fcc Brillouin zone. 
   Weyl-Kondo semimetal with $V=7.5$ and nodes pinned at $E_F$.
   This develops from Fig.~\ref{fig:still-trivial}(c) when the strong coupling is renormalized $\tilde{V}\rightarrow rV>0$.
The bandwidth of the upper quartet of bands is approximately that of the conduction electron bandwidth $\sim D$, 
whereas the strongly-renormalized lower quartet of heavy bands corresponds to the $k_BT_K$ energy scale.
    The parameters are $(E_d,\ell,r,V,\lambda,m)\simeq(-7,7.279,0.220,7.5,0.5,1)$.} 
\end{figure}

We now wish to consider our model when the hybridization is nonzero, so that the conduction $c$ and strongly correlated $f$ electrons are coupled with each other.
The choice of chemical potential and energy level $E_d=E_F$ before turning on hybridization is arbitrary, and was 
made
 to adiabatically connect the trivial insulator phase shown in Fig.~\ref{fig:still-trivial} to the Kondo regime.

In the course of solving the saddle-point equations self-consistently, the starting value for $E_d$ is far below $E_F$, 
and $\mu$ is determined analytically from the eigenergies as $\mu=-\frac{(rV)^2}{E_d+\ell}$ (see Appendix~\ref{app:pmD}).
A properly ``strong'' coupling solution usually means that $r$ is small but nonzero, which arises for a range of $V$ larger than some critical value.
The small bosonic field measures a small but nonzero hole fluctuation $r^2$ away from $n_d=n_f=1$, which is only coupled to $V$.
A valid self-consistency solution always finds an $r,~\ell$ that fixes the densities to the values specified.
The solutions at this filling fix the Fermi energy such that of 8 total bands, 2 Kondo-driven bands are filled, which corresponds to the $\nu=2n$ filling enforcement condition.

A solution of this type was shown in Ref.~\onlinecite{WKSM_PNAS}. 
To demonstrate its robustness, we solve the case with a different set of parameters. 
The result is plotted in Fig.~\ref{fig:kondopin}, which shows the conduction electron bands with bandwidth $D$ unoccupied, well-separated by a gap of $\sim$$6t$ from the renormalized narrow $f$-bands with heavy Weyl-Kondo quasiparticle excitations around nodes fixed precisely at $E_F$.  
This demonstrates that the nodal states develop out of Kondo effect.
The Kondo effect correlations produce this topological phase transition from trivial band insulator to WKSM, and \textit{pin} the nodes to the Fermi energy as a fundamental property.

\begin{figure}[t]
   		\includegraphics[width=\columnwidth]{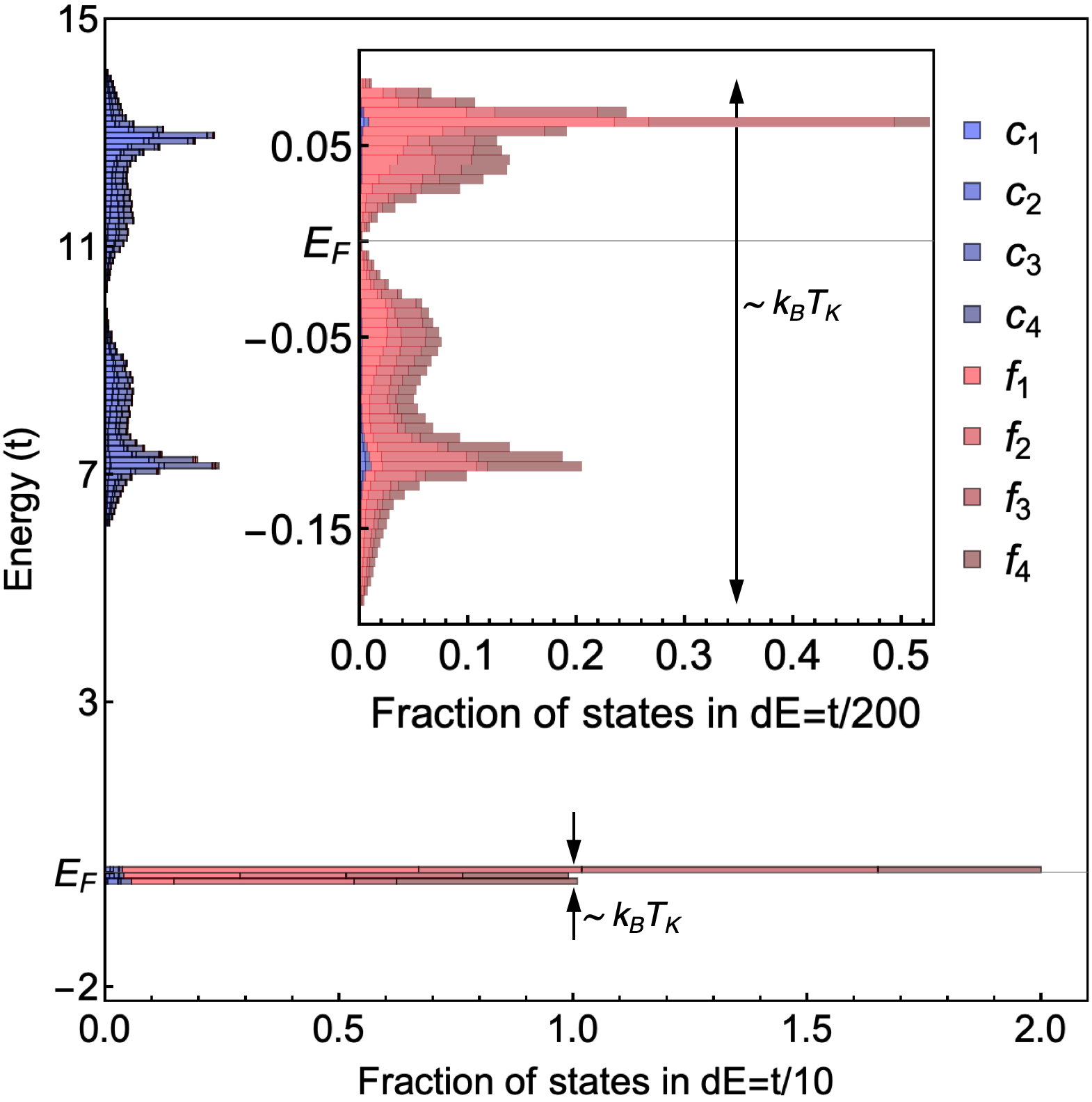}
   \caption{ \label{fig:qf_dos_w0} 
   Projected density of states, showing full energy range corresponding to Fig.~\ref{fig:kondopin}.
    Inset: zoomed-in to localized set of bands near the nodes at $E_F$. 
   Shades of red indicate contributions from $f$-fermions, and shades of blue indicate contributions from $c$-electrons.
   The parameters are $(E_d,\ell,r,V,\lambda,m)\simeq(-7,7.279,0.220,7.5,0.5,1)$.}
\end{figure}

The role of the localized species near the Fermi energy can also be demonstrated by calculating the projected density of states. 
This is provided in Fig.~\ref{fig:qf_dos_w0}, which corresponds to the hybridized parameters generating Fig.~\ref{fig:kondopin}.
The contributions of the $c$ and $f$ fermions are represented in shades of blue and red, respectively.
The main panel of Fig.~\ref{fig:qf_dos_w0} shows that at the Fermi energy, the proportion of localized $f$ fermions is large compared to that of the conduction $c$ electrons, using an energy interval of $dE=t/10$.
The inset of Fig.~\ref{fig:qf_dos_w0} shows a zoomed-in view of the projected density of states with a smaller energy interval of $dE=0.005t$ to accommodate the reduced bandwidth.
In the energies closest to $E_F$, the $f$ fermions prominently characterize the states compared to the $c$'s.
The hybridization has allowed a tiny amount of $c$-electrons to mix (see inset) through the hole fluctuations of the $r$-bosonic condensate via $n_c=r^2$.
This demonstrates that the localized $f$ electrons are directly responsible for producing the Weyl-Kondo semimetal.

We close this subsection with two remarks.
First, other quantities can also be calculated. 
For example, the surface states that correspond to the same parameter choice has already been shown in Fig.~\ref{fig:surfacestates}.

Second, going beyond the saddle-point level, the renormalized quasiparticles will acquire a finite lifetime due to the residual interactions. 
However, because the Kondo-driven Weyl nodes are pinned at the Fermi energy, the strongly renormalized nodal excitations will be long-lived, with the lifetime reaching infinity when the node is approached. 
This makes the Kondo-driven Weyl nodal excitations well defined even though it is a strongly interacting many-body system.

\section{\label{sec:tilt} Tilted Weyl-Kondo semimetal}

In the process of understanding a large spontaneous Hall effect observed in \cepd,~\cite{2018arXiv181102819D} a tilted variation of the Weyl-Kondo solution was introduced there. 
Here, we further investigate this effect. 
This allows us to analyze the details of the Berry curvature distribution near a small Fermi pocket centered around the Weyl nodes, and how this distribution can be made extremely asymmetric 
with respect to
a Weyl or an anti-Weyl node by the tilting potential.
Our results further support the analysis presented in Ref.~\onlinecite{2018arXiv181102819D}.

\begin{figure}[t]
   \centering	
      	\quad\includegraphics[width=.5\columnwidth]{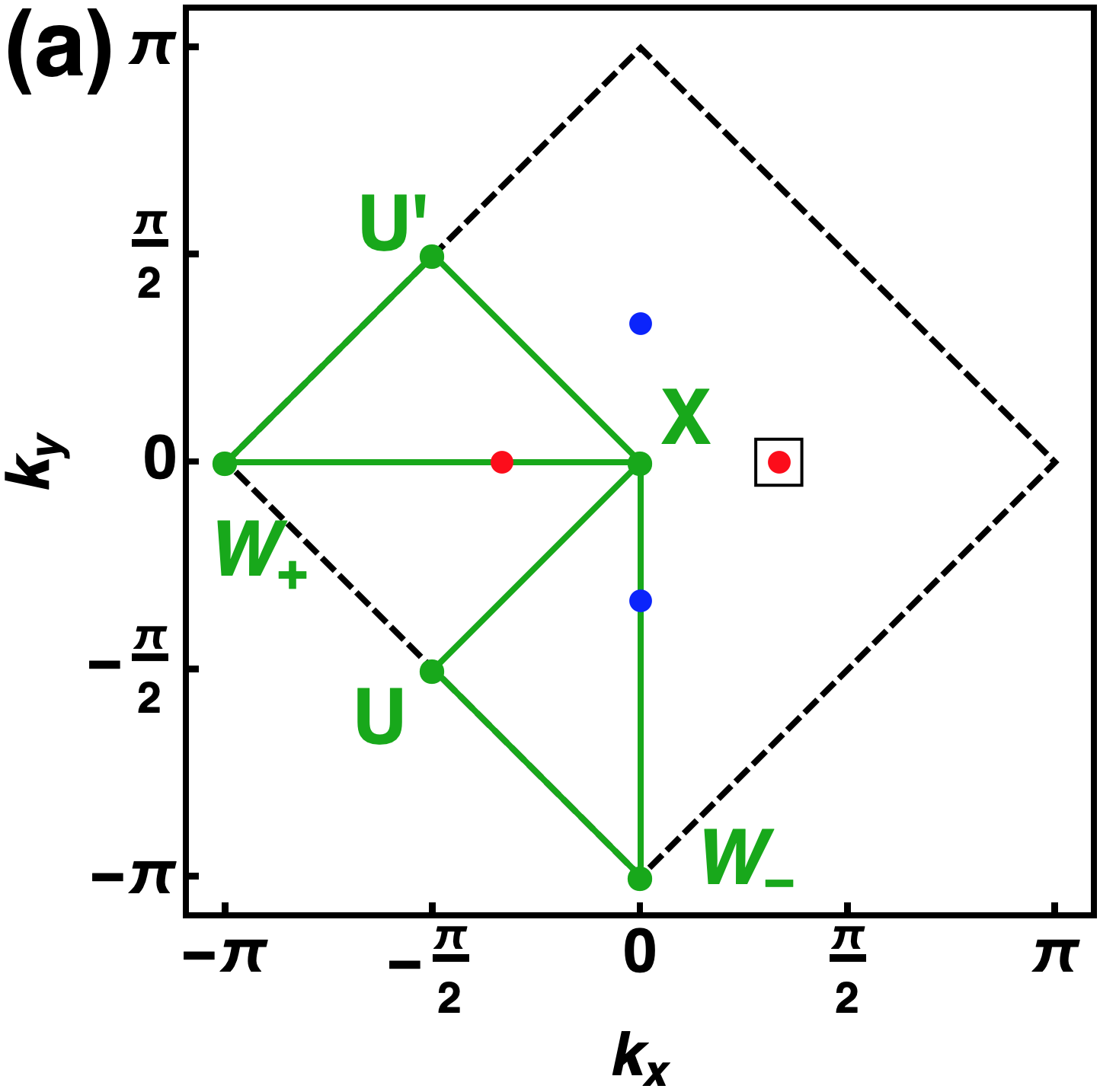}\\	
		\includegraphics[width=.49\columnwidth]{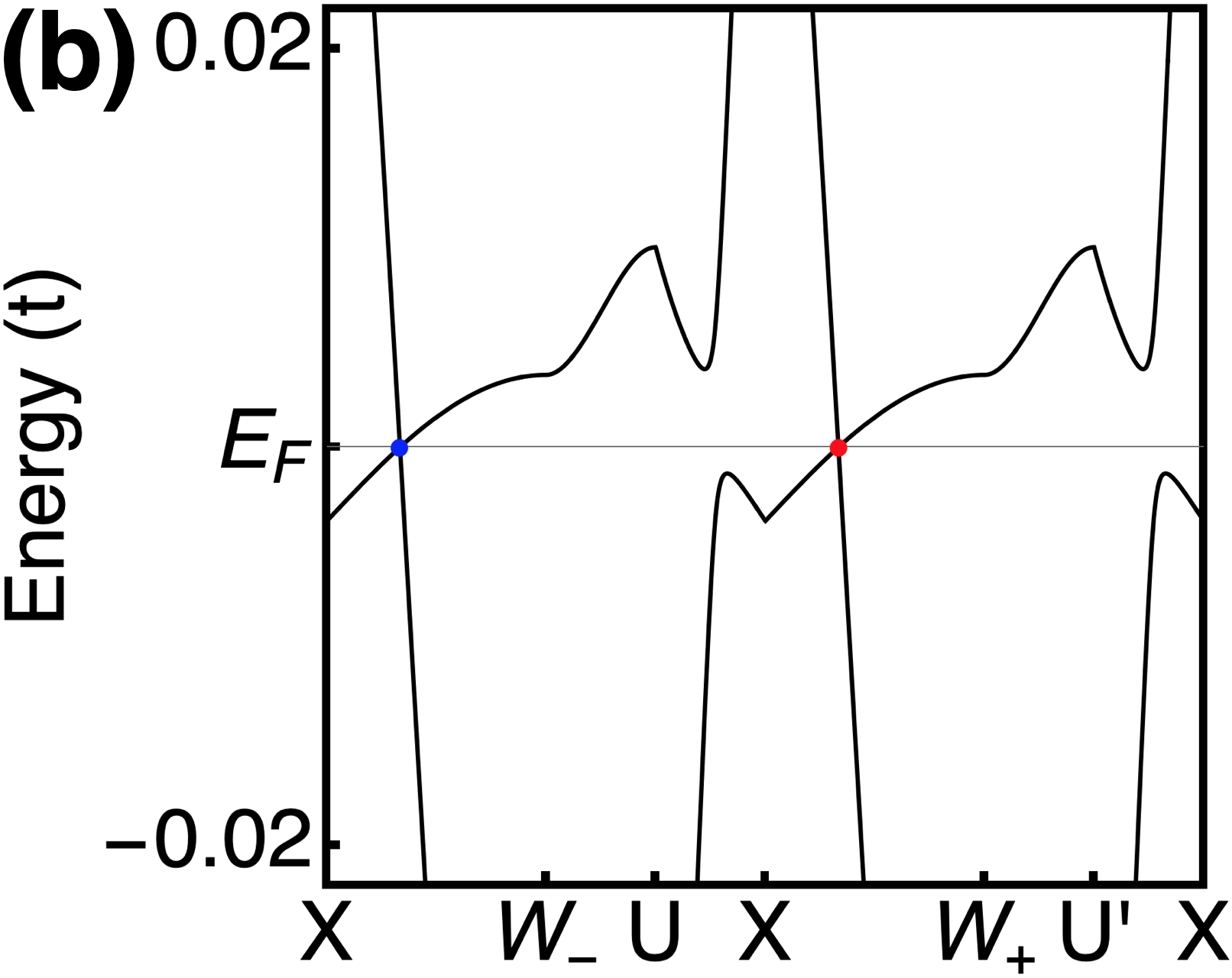}	
   		\includegraphics[width=.49\columnwidth]{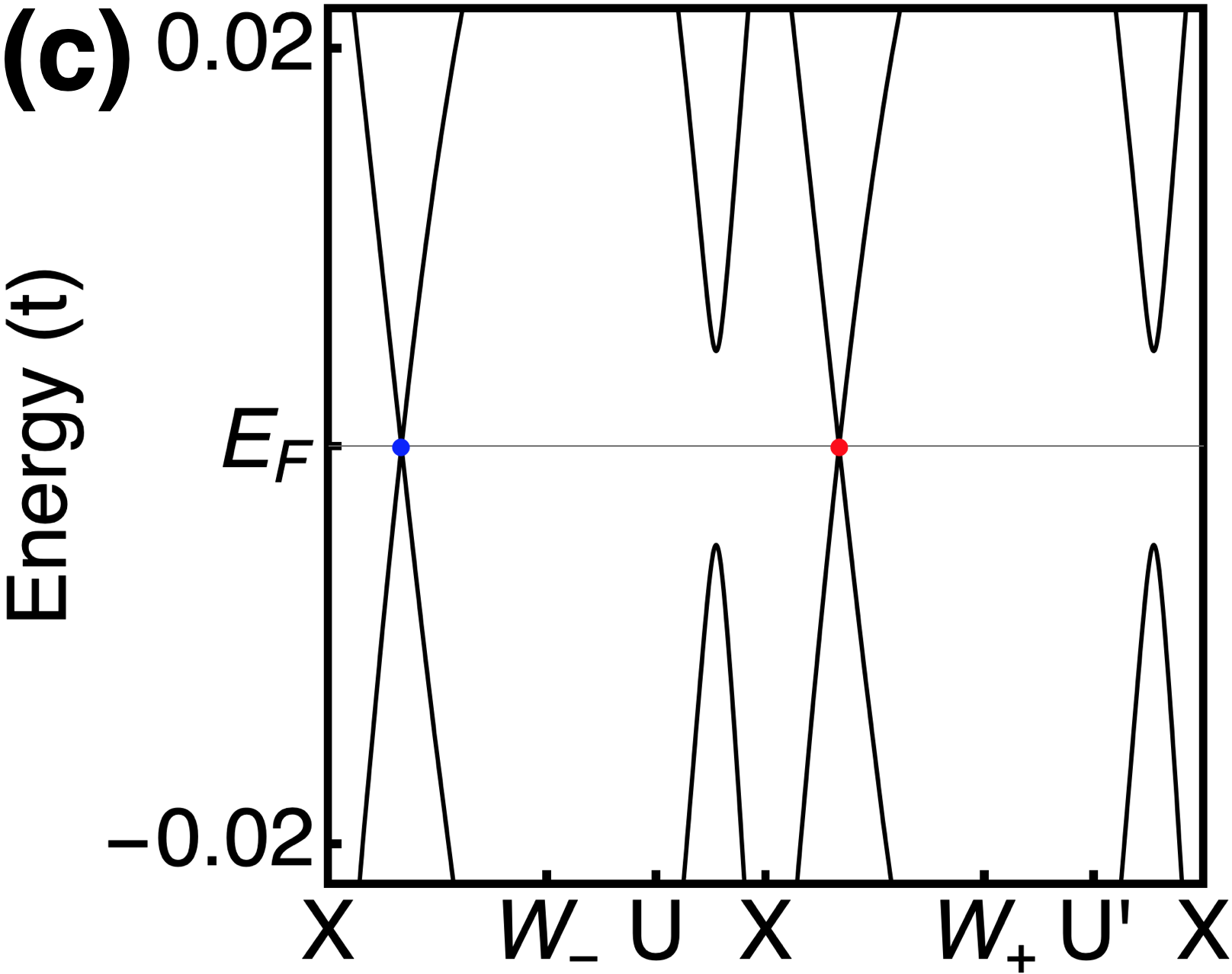}
   \caption{ \label{fig:tilt_disp} 
   \textit{High-symmetry dispersion of the tilted WKSM model.}
   (a) high symmetry contour in green along the $\k=(k_x,k_y,2\pi)$ Brillouin zone boundary plane, 
   through (anti)nodes marked in (red)blue.  
    The square marks the small region of the Brillouin zone over which the Berry-curvature distribution is shown in Fig.~\ref{fig:berry}(a).
   (b) Bulk dispersion along the contour of the tilted model with $C=0.8$, other parameters are found self-consistently to be $(E_d,\ell,r,V,\lambda,m)\simeq(-7,7.282,0.222,7.5,0.5,1)$.
   (c) Bulk dispersion with $C=0$, and the same parameters Fig.~\ref{fig:kondopin}.
      }
\end{figure}

\subsection{\label{sec:tilted} Tilted Weyl dispersion }

The tilting term in our diamond-lattice model~\cite{2018arXiv181102819D} of the WKSM can locally adjust the anisotropy of the linear part of the dispersion.
We specify it as 
\begin{align}
	\mathcal{H}_t &= C\sum_{\k,\sigma}\left[1-\frac{1}{2}D(\k)\right]n^c_{\k \sigma},\label{eq:tilt}
\end{align}
which preserves the lattice symmetry and is added to the conduction electron Hamiltonian Eq.~(\ref{eq:condel}). 
Here, $C$ sets the tilting potential, and we continue to consider the specific parameters of $m=1$ and $\lambda=1/2$.
We have solved the saddle-point equations in the presence of the tilting term.

The resulting dispersion is contrasted to that of the un-tilted model, as shown in Fig.~\ref{fig:tilt_disp}.
The Weyl nodes remain pinned at the Fermi energy in the tilted case. 
This further illustrates the robustness of the mechanism discussed in the previous section for the formation and pinning of the Kondo-driven Weyl nodes.

To see the reason that the $C$ term tilts the dispersion near the nodes, note that near the Weyl node, Eq.~(\ref{eq:hminus}) is a $2\times 2 $ Hamiltonian matrix and can be linearized to obtain a $\k \cdot \bm{\tau}$ form.
The Hamiltonian $\mathcal{H}_t$ is proportional to the matrix $\tau_0\otimes\sigma_0$, which commutes with the canonical transformation, and after the transformation it contributes a term $C[1-\tfrac{1}{2}D(\k)]\tau_0$ to $h_{\k-}$.
Linearizing the full $h_{\k-}$ near the nodes gives a linear dispersion which adds velocity components that depend on the tilting direction $\hat{t}$ as $\v_t=C\hat{t}$.
Now the effective Hamiltonian is 
\begin{align}
	H_{\text{eff}} &=\v_t\cdot\k \tau_0 + v \k \cdot \btau.
\end{align} 
Using the velocity ratio, the effective Hamiltonian has the regimes $|\frac{C}{v}|<1$ (type I), 
and $|\frac{C}{v}|>1$ (type II).~\cite{Soluyanov:2015aa}
The type-I behavior makes the dispersion anisotropic, and causes the Fermi surface to change shape within the BZ.
A type-II Weyl semimetal arises when the tilt has become extreme enough to cause a Lifshitz transition of the Fermi surface.

In Fig.~\ref{fig:tilt_disp} we illustrate the tilting of the linear bands around the node when $C\neq0$; we plot the eigenenergies along the green high-symmetry $\k$-contour in Fig.~\ref{fig:tilt_disp}(a) that intersects with the Weyl nodes.
Fig.~\ref{fig:tilt_disp}(b) shows the dispersion of a strong coupling limit solution when $C=0.8$, where there are anisotropic slow and fast bands along $X-W_\pm$.
There also appears to be two Lifshitz transitions ready to happen: the type-I to type-II tilting transition, and the Fermi pocket lowering itself to the Fermi energy around $U$ and $U'$.
This is in contrast to the non-tilted $C=0$ dispersion shown in Fig.~\ref{fig:tilt_disp}(c), where the linear part of the dispersion appears isotropic along $X-W_\pm$ for energies sufficiently close to $E_F$.

	\subsection{\label{sec:bc} Berry curvature distribution}

\begin{figure}[t]
	\centering
   		\includegraphics[width=0.7\columnwidth]{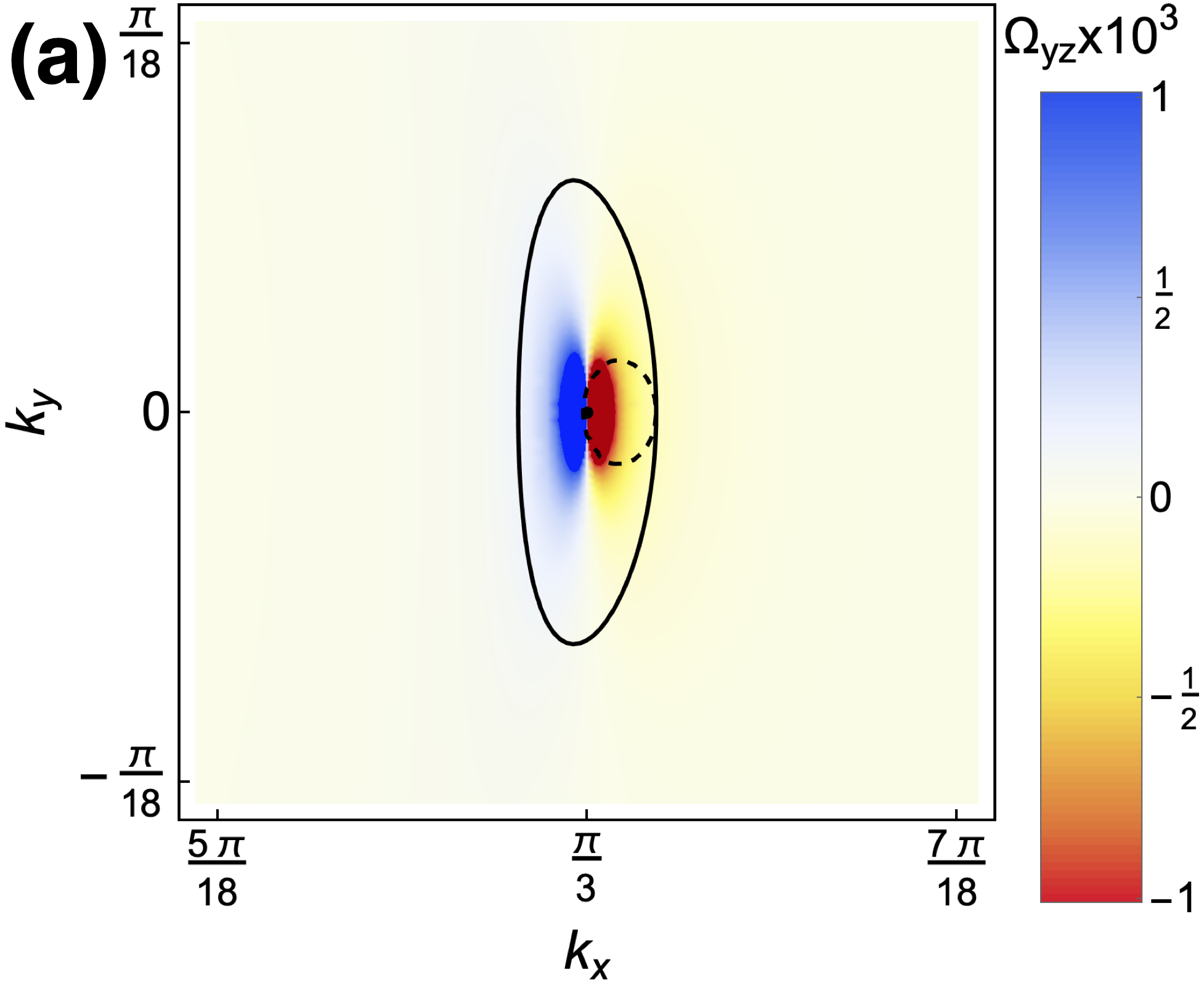}\\
		\includegraphics[width=0.72\columnwidth]{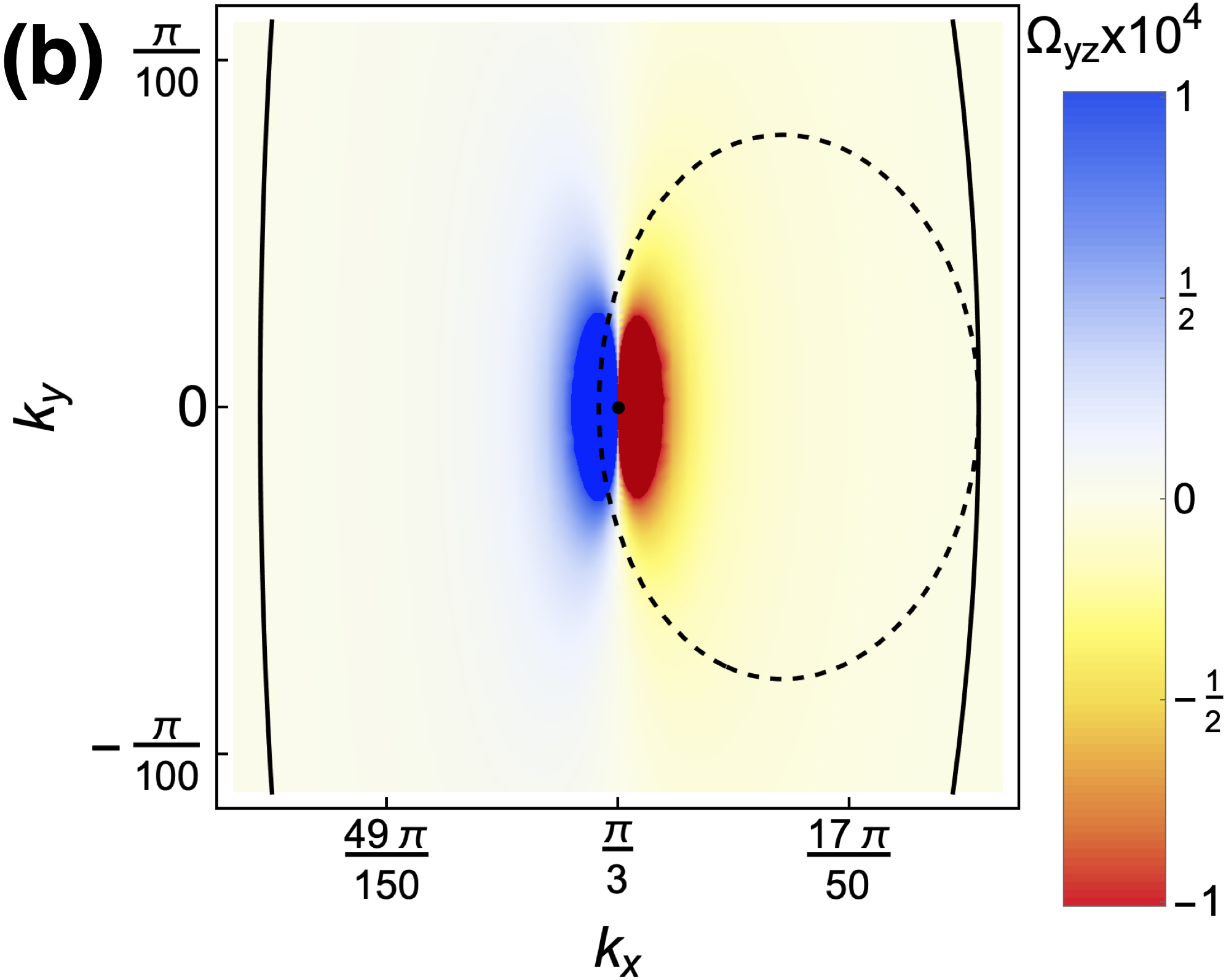} 
	\caption{ \label{fig:berry} 
	\textit{ Berry curvature strength captured by Fermi surfaces.} 
	The Berry curvature component $\Omega_{yz}$ [Eq.~(\ref{eq:bc})] of the WKSM phase in the $[001]$ plane of the Weyl node located at the center of the plot, $\k=(\tfrac{\pi}{3},0,2\pi)$ (black dot), which corresponds to the right red point of Fig.~\ref{fig:tilt_disp}(a), and Fermi surfaces from a slightly metallic filling $n_d+n_c=1+10^{-5}$, sharing the parameters $(E_d,V,\lambda,m)\simeq(-7,7.5,0.5,1)$.
	(a) The solid contour is the intersection of the three dimensional Fermi surface of the model with the BZ-boundary plane ({\it i.e.}, for $k_z=2\pi$) without tilting, and self-consistently determined parameters 
	$(C,\mu,r,\ell)\simeq(0,-9.748,0.220,7.279)$.
	The dashed contour is the counterpart for the Fermi surface of the tilted model, with parameters 
	$(C,\mu,r,\ell)\simeq(0.9,-9.811,0.222,7.281)$.
	(b) Same as (a), but with an even smaller plot range in $\k$-space.
	The full range of the Berry curvature is truncated, and the deep blue and red regions near the node represent values that extend beyond the legend.
	Note that, in the three dimensional BZ, the Fermi pockets in the two cases have the same volume.
}
\end{figure}

The physical quantity that lies at the heart of an electronic topological phase's ``topologicalness'' is the Berry phase, and we show in Fig.~\ref{fig:berry} the Berry curvature distribution for a tiny portion of the BZ that surrounds 
one of the Weyl nodes of Fig.~\ref{fig:tilt_disp}(a) (denoted by the small square surrounding the right red node of Fig.~\ref{fig:tilt_disp}(a)).
The Berry phase is the condensed matter manifestation of the geometric phase acquired when a wave system explores the landscape of its parameter space as a result of a cyclic adiabatic process.
In condensed matter systems, a particle in state $n$ acquires a Berry phase $\gamma_n$ as the Hamiltonian parameters are varied around a closed path $\mathcal{S}$ without eigenstate transitions away from $n$ ({\it i.e.}, adiabaticity is preserved).
The Berry phase and the related Berry flux quantity is expressed in terms of the Berry vector potential $\mathcal{A}_n(\k)$ and corresponding Berry curvature $\Om_n(\k)$ as
\begin{align}
	\Phi_n(\mathcal{S})&=\oint_{\mathcal{S}} d\k \cdot \mathcal{A}_n(\k)\label{eq:bf1}\\
			&=\frac{1}{2\pi}\int_{\partial\mathcal{S}} d\S \cdot \Om_n(\k)\\
			&=\frac{1}{4\pi}\int \frac{d^3k}{(2\pi)^3} \nabla_\k\cdot\Om_n(\k),\label{eq:bf2}
\end{align}	
where		
\begin{align}			
	\Om_n(\k)&=\nabla_\k\times\mathcal{A}_n(\k)\\
			&=\left(\Omega_n^{yz}(\k),\Omega_n^{zx}(\k),\Omega_n^{xy}(\k)\right),\label{eq:bcvec}\\
	\Omega^{ab}_n(\k)&=\sum_{n\neq n'}
	\text{Im}\frac{ \la n \k|\partial_{k_a} H_\k|n' \k\ra\la n' \k|\partial_{k_b} H_\k|n \k\ra }{(\mathcal{E}_n - \mathcal{E}_{n'})^2},\label{eq:bc}
\end{align}
One immediately sees that at a Weyl node degeneracy, the denominator of Eq.~(\ref{eq:bc}) is zero, causing a singularity in the Berry curvature field in the momentum space [Eq.~(\ref{eq:bcvec})].
Due to their singular structure, the positive and negative Berry curvature singularities corresponding to Weyl nodes and anti-nodes possess a local topological invariant that can distinguish between nodal signs, and is just the Berry flux of Eq.~(\ref{eq:bf2}) computed around a single node.
For a single Weyl node, the Berry flux density is $\nabla\cdot\Om=\pm4\pi\delta^3(\k)$ and the Berry flux is
\begin{align*}	
		\Phi(\mathcal{S})&=\pm\frac{1}{4\pi}\int \frac{d^3k}{(2\pi)^3} 4\pi\delta^3(\k)
		=\pm 1.
\end{align*}
This tells us that the Berry flux is a quantized number that counts the number and sign of Berry charges (monopoles) in a given system.
This is the reason for the ``monopole'' terminology: the Berry flux acts like the total charge of the Weyl point, and their associated Berry curvature (magnetic) field has a monopole or anti-monopole configuration.
When the Weyl (anti-)nodes are pinned to the immediate vicinity of the Fermi energy, as happens in our Weyl-Kondo semimetal solution,  the monopoles and associated Berry curvature singularities appear very close to the Fermi surface.
In other words, the Fermi surface comprises tiny Fermi pockets that surround the Weyl (anti)nodes, and states on the Fermi surface have a very large Berry curvature.

An intriguing question is how to tune the singular nature of the Berry curvature. 
Tilting the Weyl cone dispersion is one means of doing so. 
An example of the tilt behavior of the model is seen in Fig.~\ref{fig:berry}.
The color scale indicates the magnitude of the $\Omega_{yz}$ component of the Berry curvature, 
which is highly concentrated around the node and discontinuous at the node, located in the center of each plot.
The solid and dashed contours show the Fermi surface produced with $C=0$ and $C=0.9$, respectively, 
from a slightly metallic filling $n_d+n_c=1+10^{-5}$.
For both cases, but especially for the tilted case, the Berry curvature is very large on the Fermi surface, reflecting the proximity of the Fermi surface to a Weyl monopole.

In Fig.~\ref{fig:berry}(a), the area in the momentum space shown corresponds to the small square surrounding the right red node of Fig.~\ref{fig:tilt_disp}(a).
This conveys how tiny the Fermi surfaces are compared to the extent of the BZ.
In Fig.~\ref{fig:berry}(b), we zoom in further from Fig.~\ref{fig:berry}(a) to the vicinity of the node.
Here, it is apparent that the $C=0.9$ tilted Fermi surface (dashed line) still encloses the node, 
but it is much closer to the node as compared to the $C=0$ Fermi surface (solid line).
The tilting term is seen to make the Fermi pocket and the associated Berry curvature be distributed 
around the Weyl node in a highly asymmetrical way,
in which highly singular Berry curvature fields strongly influence the heavy Weyl quasiparticles pinned at the Fermi surface.

In particular, Figs.~\ref{fig:berry}(a) and \ref{fig:berry}(b)
illustrate that the asymmetry induced by the tilting term has made one side of the Fermi pocket to be much
closer to the node than the other. 
This makes an external electric field to be readily able to drive the system to a highly out-of-equilibrium response in the 
Berry-curvature-induced transverse conductance, a theoretical framework that was advanced for the spontaneous Hall effect in Ref.~\onlinecite{2018arXiv181102819D}.
As such, our result concretely demonstrates the ready realization of the giant spontaneous Hall effect put forward there.

\section{\label{sec:sigs} Signatures of correlated topological semimetals} 

Some of the conventional signatures of Weyl semimetals are 
quantum oscillation experiments, negative longitudinal magnetoresistance indicative of the chiral anomaly, and angle-resolved photoemission spectroscopy (ARPES) imaging of both the bulk and surface states, which probes the linear nodal dispersion in the bulk and verifies topology due to the bulk-boundary-correspondence.
These experiments elegantly illustrate the signatures of \textit{weakly correlated} Weyl semimetal material candidates.

As has been pointed out recently,~\cite{2018arXiv181102819D} the \textit{strongly correlated} Weyl-Kondo semimetal phase offers several obstacles to characterization by conventional experimental probes.
For example, imaging the surface states would have to use an ARPES setup that could resolve features within a bandwidth of approximately $D^* \sim k_BT_K$, with for example, in \cepd, 
$T_K\approx13\,$K, which would require an ultrahigh resolution much
below $D^* \sim1\,$meV.

On the other hand, the WKSM displays unique physics that offers more suitable probes.
The WKSM phase exhibits node pinning that is contingent on the development of the Kondo effect, a slow effective Weyl fermion velocity, and, correspondingly, a narrowed bandwidth.
In Refs.~\onlinecite{WKSM_PNAS,DzsaberPRL2017}, it was shown that the specific heat has the following form:
\begin{eqnarray}
	C_p = \Gamma \,T^3 \,.
\end{eqnarray}
The prefactor of the $T$-cubic dependence from the contribution of each node
is
\begin{eqnarray}
	\Gamma =\frac{7\pi^2}{30}k_B\left(\frac{k_B}{\hbar v^*}\right)^3 \,.
\end{eqnarray}
This means that $\Gamma$ is enhanced relative to the typical non-interacting value by a factor of $(v/v^*)^3$, the ratio of the typical Fermi velocity of a non-interacting electron band to the renormalized velocity of the Kondo-driven Weyl nodal excitations; 
this enhancement is robust against the residual interactions between the Weyl fermions.
The $v^*$ value extracted for \cepd \, is $\sim$$886\,$m/s, three orders of magnitude smaller compared to a normal metal's Fermi velocity.~\cite{DzsaberPRL2017}

The Kondo-driven node pinning also implies that a large Berry curvature singularity at the Fermi energy develops at $T<T_K$.
This provides a means of using the Berry-curvature-induced anomalous velocity 
to probe the topological characteristics of the Kondo-driven Weyl nodes in the WKSM (that is time-reversal invariant in equilibrium), at zero magnetic field and in a nonlinear response to an applied electric field,
as has recently been demonstrated 
in Ref.~\onlinecite{2018arXiv181102819D}.

\section{\label{sec:conclusion} Conclusions and Outlook}

We have expanded on several theoretical aspects of the Weyl-Kondo semimetal state in a noncentrosymmetric Kondo/Anderson lattice model with both strong correlations and large spin-orbit coupling.
This state was advanced concurrently in theoretical \cite{WKSM_PNAS} and 
experimental \cite{DzsaberPRL2017,2018arXiv181102819D} studies. 
It preserves the time-reversal invariance. 
The Weyl nodes are driven by the Kondo effect and, 
thus, must appear within the narrow energy range near the Fermi energy for the Kondo resonance;
at the same time, their existence can be traced to the degeneracy of electronic states enforced 
by nonsymmorphic space-group symmetry. 
These two features combine to pin the Weyl nodes to the immediate vicinity of the Fermi energy. 
Moreover, the Kondo-driven nature makes the linearly-dispersing Weyl nodal excitations to have an energy scale $k_BT_K$, which is smaller than the bare conduction-electron bandwidth $D$ by orders of magnitude ({\it c.f.} Fig.~\ref{fig:kondopin}). 
Correspondingly, the velocity $v^*$ is reduced from typical values of noninteracting electrons by several orders of magnitude. 
An immediate consequence of such a reduced velocity is that the specific heat $c_V=\Gamma \,T^3$, with the $T$-cubic prefactor $\Gamma$ enhanced from the typical non-interacting value by a huge factor of $(v/v^*)^3$.
The pinning of the Weyl nodes to the immediate vicinity of the Fermi energy also implies that the Berry curvature singularities of the Weyl nodes appear near the Fermi energy. 
This gives rise to a large anomalous velocity for the states on a small Fermi surface pocket surrounding the Weyl nodes. 
As such, this pinning of the Berry curvature singularities near the Fermi energy presents a means of probing topological characteristics of Weyl nodal excitations through a spontaneous Hall effect, which is a nonlinear response to an applied electric field, even though the system under equilibrium preserves time-reversal symmetry.

The experimental developments have taken place in the Kondo-driven semimetal \cepd, 
a cubic system for which the space group is nonsymmorphic (no. 220), the inversion symmetry is broken, but the time-reversal symmetry is preserved.~\cite{2018arXiv181102819D}
This new heavy fermion semimetal shows a $T^3$ specific heat with a huge prefactor $\Gamma$, so much so that it surpasses the phonon contribution,~\cite{DzsaberPRL2017} and a giant spontaneous (zero magnetic field) Hall effect and an accompanying even-in-magnetic-field component.~\cite{2018arXiv181102819D} 
The results provide direct evidence for ultraslow Weyl nodal excitations and its topological nature.

We close with a look into future directions.
First, the developments along this direction point to the search for further Weyl-Kondo semimetals in heavy fermion systems with nonsymmorphic space groups, as already exemplified by the case of \cepd.~\cite{DzsaberPRL2017,2018arXiv181102819D,dzsaber2019quenching}
Because the majority of the 230 space groups in three dimensions are nonsymmorphic, this suggests the prevalence of Weyl-Kondo semimetal phases in such systems.
Of potential interest in this context include nonsymmorphic heavy fermion semimetals with broken inversion symmetry such as
CeRu$_4$Sn$_6$~\cite{Paschen2010,Guritanu2013,Sunderman2015,Wissgott2016,Yu2016} and CeNiSn,~\cite{StrigariThesis2015,Sto16.1,Bareille2019} and those that are inversion symmetric but with time-reversal symmetry broken
by an external magnetic field or magnetic ordering, such as 
YbBiPt~\cite{Fisk1991,Chadov:2010aa,Mun-2013,Guo:2018aa} and CeSbTe.~\cite{Schoopeaar2317,Schoop2018,Topp2019}

Second, we have stressed that the Weyl-Kondo semimetal solution is robust because of the cooperation of the Kondo effect with the space-group symmetry. It will be instructive to explore the role of space-group symmetry on Kondo-driven bulk nodal excitations near the Fermi energy in related models.~\cite{Feng13,Feng2016,Pixley2017,Cook2017,Dzero2016,Chang:2017aa,PhysRevB.89.085110}

Third, the considerations of symmetry open up different ways to think of accessing nearby topological phases, by reducing or restoring point-group or space-group symmetries.
One path that has been explored is anisotropically tuning the hopping amplitudes $t_{ij}$ of the lattice bonds.~\cite{FKMmodel07,PhysRevB.78.165313} 
Such a symmetry-reducing tuning could be approximated through uniaxial stress, which, in Kondo systems, also tunes the strength of the Kondo effect.
Another avenue is to explore nonspatial symmetries, such as time-reversal symmetry breaking.~\cite{dzsaber2019quenching}
In a one fermion flavor model,~\cite{grefe2020jps} we have established that a tunable TRSB term can coexist with the Weyl semimetal phase described in Sec.~\ref{sec:realization}, but that when the TRSB term is larger than the ISB term, a topologically distinct Weyl semimetal phase can emerge with nodes in the Brillouin zone interior.
It is an exciting next step to incorporate TRSB to the full Kondo-driven model.
Finally, doping studies represent a promising way of tuning.~\cite{DzsaberPRL2017,Cao2019arxiv}

Fourth, the theoretical and experimental results on the Weyl-Kondo semimetal sets the stage to address how the overall quantum phase diagram of heavy fermion metals, Fig.~\ref{fig:qpt-hf}(b), is enriched by topologically nontrivial metallic phases driven by the combined effects of strong correlations and spin-orbit coupling. 
A recurring theme of heavy fermion metals is that novel phases develop in the quantum critical regime, at the border of electronic orders. 
This reflects the accumulation of entropy in the quantum critical 
regime,~\cite{si_PRL_grun_03,Wu_2011,Rost.09,Luc17} 
as a result of which the electronic matter is soft and prone to developing novel phases. 
A canonical example of such emergent phases is unconventional superconductivity,~\cite{Mathur98,Si_PSSB10} 
but it could also be nematic or other forms of secondary electronic orders.~\cite{Ron17.1} 
An intriguing possibility is that, when the spin-orbit coupling and correlations are both strong, topologically nontrivial metallic states appear as emergent phases at the border of electronic order, albeit on the nonordered 
[ i.e., disordered; cf. Fig.~\ref{fig:qpt-hf}(a)]
side. 

Finally, the approach taken here represents a general means of treating the space-group symmetry enforcement 
of topological semimetals in strongly correlated settings. 
In the theoretical model, the Weyl nodes of the bare conduction electrons enforced by the space group symmetry 
are located far away from the Fermi energy. When the Kondo effect takes place, the Weyl nodes 
are transmitted to those of the Kondo-driven quasiparticles. 
The combination of strong correlations and space-group symmetry enforcement pins the Kondo-driven Weyl nodes to the immediate vicinity of the Fermi energy.
This makes the strongly correlated Weyl-nodal excitations to be well-defined, even for a underlying many-body system
that is strongly interacting. Equally important, it allows the theory to connect with the striking experiments 
in \cepd.~\cite{DzsaberPRL2017,2018arXiv181102819D,dzsaber2019quenching}
This type of interplay between the space-group symmetry constraint and strong correlations 
is likely to be important in other settings of strongly correlated topological matter as well.

\begin{acknowledgements}
We thank Jennifer Cano, Sami Dzsaber, 
Leslie Schoop, Steffen Wirth, and Jianxin Zhu for illuminating discussions.
The work at Rice has been supported by the NSF (DMR-1920740), the Robert A. Welch Foundation (C-1411) and the ARO (W911NF-14-1-0525). 
Work in Vienna was supported by the Austrian Science Fund (DK W1243, I2535, I4047, and P29296).
\end{acknowledgements}


\appendix

\section{ \label{app:pmD} 
Existence of the Weyl-Kondo semimetal nodes}

Here we show that in our model, the eigenenergies have Weyl nodes at the Fermi energy.
First, the method for obtaining the eigenenergies was detailed in Ref.~\onlinecite{WKSM_PNAS}, and the eigenenergies are
\begin{align}
	\mathcal{E}^{(\tau,\alpha)}_{\pm D}(\k)&= \frac{1}{2} \left[ E_s + \tilde{\eps}^{\tau}_{\pm D}  +\alpha \sqrt{\left( E_s - \tilde{\eps}^{\tau}_{\pm D}  \right)^2 + 4V_s^2}\right],\label{eq:disp}\\
	\tilde{\eps}^{\tau}_{\pm D} &=\eps^{\tau}_{\pm D}-\mu, \label{eq:c-disp-mu} \\
	\eps^{\tau}_{\pm D} &=\tau\sqrt{u_1^2(\k)+u_2^2(\k)+(m\pm \lambda D(\k))^2}, \label{eq:c-disp} \\
	u_1(\k)&= t\left(1+\sum_{n=1}^3\cos(\k\cdot {\bf a}_n)\right),\label{eq:u1}\\
	u_2(\k)&= t\sum_{n=1}^3 \sin(\k\cdot{\bf a}_n ),\label{eq:u2}\\
	D(\k)&=2\left\{ \sin^2\left(\tfrac{k_x}{2}\right) 
	\left[\cos \left(\tfrac{k_y}{2}\right)-\cos\left(\tfrac{k_z}{2}\right)\right]^2\right.\nonumber\\
	&+\sin^2\left(\tfrac{k_y}{2}\right) \left[\cos \left(\tfrac{k_x}{2}\right)-\cos \left(\tfrac{k_z}{2}\right)\right]^2\nonumber\\
	&\left.+\sin^2\left(\tfrac{k_z}{2}\right) \left[\cos \left(\tfrac{k_x}{2}\right)-\cos \left(\tfrac{k_y}{2}\right)\right]^2 \right\}^{1/2},\label{eq:Dkfull}
\end{align}
where the index which labels $\tau=\pm1$, $\alpha=\pm1$, $\pm D=\pm D(\k)$ distinguishes the eight bands of the system, and the ${\bf a}_n$ are the primitive lattice vectors of the diamond lattice.
The Hamiltonian is only separable in terms of the $|\pm D\ra$ pseudospin basis, and $\alpha$ distinguishes between the upper and lower quartet of bands; 
within each quartet, $\tau$ distinguishes the upper two from the lower two bands, and one can write
\begin{align}
 	\tilde{\eps}^{\tau}_{\pm D} &=\tau\eps_{\pm D}-\mu,\nonumber\\
	\eps_{\pm D} &=\sqrt{u_1^2(\k)+u_2^2(\k)+(m\pm \lambda D(\k))^2}.\label{eq:c-disp-tau} 
\end{align}
Without loss of generality, we take the parameters $t,r,V\geq0$.
There are line or Dirac nodal touchings in the case of $m=0$, depending on whether $\lambda=0$ or nonzero.
Here, we are only interested in Weyl node touchings, and thus we consider only the case with both $\lambda,m$ being nonzero; for definiteness, we focus on $\lambda,m>0$.

The periodic Anderson model also has a few additional specifications.
In the Kondo regime, we have
\begin{align}
	E_s &= E_d+\ell,\nonumber\\
	V_s &= rV, \label{eq:strongweak}
\end{align}
and $V>V_c$ for some critical value, beyond which the $r$-bosonic field is small but nonzero.
It is taken that the bare localized fermions representing the $4f$ electrons should have an energy level $E_d$ that has a bare value well below the conduction-electron bands,
$E_d\ll \eps^\tau_{\pm D}$.
We chose to define $E_F=0$ and $E_d<0$; we will determine the signs of $\mu,\ell$ near the end, and in Appendix~\ref{app:mu}.

With these pieces in place, it is simple to observe the following.
Since all parameters and eigenenergies are real, 
and each term in the summation within each square root is squared (nonnegative), the square root quantities are also nonnegative [Eqs.~(\ref{eq:disp}),(\ref{eq:Dkfull}),(\ref{eq:c-disp-tau}), and Eq.~(\ref{eq:Dk})].
In turn, Eq.~(\ref{eq:c-disp-tau}) implies that
\begin{align}
	\eps_{+D}\geq\eps_{-D}\geq0,\label{eq:c-ineq}
\end{align}	
given that $m,\lambda,D(\k)>0$, since the differentiating term involving $D(\k)$ yields
\begin{align}
	|m+\lambda D(\k))|\geq m\geq|m-\lambda D(\k)|\geq0.\label{eq:pm-ineq}
\end{align}
It then follows from Eq.~(\ref{eq:c-disp-tau}) that
\begin{align}
 	-\eps_{+D} \leq -\eps_{-D} \leq 0 \leq +\eps_{-D} \leq +\eps_{+D}. \label{eq:c-disp-order} 
\end{align}

In the diamond lattice space group, a nontrivial filling enforced semimetal~\cite{Watanabe2016} occurs at filling factor $\nu=2n$, $n\neq2$. 
For a quarter of the bands to be full, 
$\nu=2(n_c+n_d)=2$ of the eight bands must be full, so the gapless band touching point should occur between the third (hole) $\mathcal{E}_3(\k)$ band, and the second (filled) $\mathcal{E}_2(\k)$ band.
We label the bands according to ascending order in energy. 
Then, the condition that determines when Weyl nodes $\k_W$ are present at quarter filling is
\begin{align}
	\lim_{\k\rightarrow\k_W}\Delta_{32}(\k)&=\lim_{\k\rightarrow\k_W}\left[\mathcal{E}_3(\k)-\mathcal{E}_2(\k)\right]=0.
	\label{eq:delta32}
\end{align}

It is clear from the form of Eqs.~(\ref{eq:disp}) and (\ref{eq:c-disp-tau})
that only the $\eps_{\pm D}$ terms are $\k$-dependent, so they shall determine where nodes may appear in momentum space.
Because of this model's underlying nonsymmorphic symmetry, we know that the nodes should appear on the BZB, and that they generically avoid three- and sixfold symmetry axes when TRS is preserved,~\cite{Kane_3ddirac} so we search on the $[001]$ face of the BZ (the other faces are related by symmetry).
We seek solutions to $\eps_{\pm D}=0$ [Eq.~(\ref{eq:c-disp-tau})].
Considering just the expression $\sqrt{u_1^2(\k)+u_2^2(\k)}=0$, this has line degeneracy solutions along the $X-W$ lines.
Therefore, the solutions to $\eps_{\pm D}=0$ will be realized for nonzero $m,\lambda$ when $\sqrt{(m\pm \lambda D(\k))^2}=0$ also, since the $|m\pm D(\k)|$ term determines the gap in the $\eps_{\pm D}$ expressions.
Along an $X-W$ line such as $\k_{XW}=(\k_x,0,2\pi)$, we already know from Eq.~(\ref{eq:pm-ineq}) that $|m+\lambda D(\k)|$ is bounded below by $m$, which we insist is nonzero.
Thus we can eliminate the $|+D\ra$ sector from consideration for the solution, and ask whether the $|-D\ra$ sector has a solution.
Along $\k_{XW}$,
\begin{align}
	\eps_{- D}(\k_{XW})&=\sqrt{(m-\lambda D(\k_{XW}))^2}\\
		&= \left|m-4\lambda\left|\sin\left(\frac{k_x}{2}\right)\right|\right|=0,\\
		&\leftrightarrow \left|\sin\left(\frac{k_x}{2}\right)\right|=\frac{m}{4\lambda},
\end{align}
so the Weyl nodes occur at
\begin{align}
	\k_W&=(k_0,0,2\pi),\\
	k_0&=2\arcsin\left(\frac{m}{4\lambda}\right),\mod\pi,
\end{align}
where the modulo $\pi$ node is the opposite chirality partner in the neighboring BZ; within the first BZ, there are six inequivalent pairs total.
Therefore we conclude that Weyl-Kondo nodes develop only for bands in the $|-D\ra$ sector.

Now it is pertinent to apply these results to the full eigenenergy expression of Eq.~(\ref{eq:disp}).
Since the sign of $\alpha$ picks out whether the band is in the upper four or lower four, we consider $\alpha=-1$ to analyze quarter filling.
Having shown that a Weyl node is permitted for the $|-D\ra$ sector, $\tau$ must be opposite in sign for each band.

Therefore, the Weyl nodes should exist between the bands $\mathcal{E}^{(-,-)}_{-D}$ and $\mathcal{E}^{(+,-)}_{-D}$.
To gain some more insight, we solve $\mathcal{E}^{(-,-)}_{-D}(\k_W)=0$, and express $\mu$ in terms of other parameters.
From Eq.~(\ref{eq:disp}), we get
\begin{align*}
	E_s -\mu- \sqrt{( E_s +\mu )^2 + 4V_s^2}&=0,\\
\end{align*}
which implies
\begin{align*}
	-E_s\mu&=V_s^2.
\end{align*}
Therefore,
\begin{align}
	\mu=-\frac{(rV)^2}{E_d+\ell},\label{eq:mu}
\end{align}
so we find that the sign of $\mu$ depends on the sign of $(E_d+\ell)$.

In our numerical calculations for the Kondo regime at quarter filling, 
the solutions always follow $(E_d+\ell)>0$;
for a given $E_d<0$, we find $\ell>0$ and $|E_d|<\ell$.
For example, in our parameter choice for Figs.~\ref{fig:kondopin}-\ref{fig:qf_dos_w0}, we had $E_d=-7$, $\ell=7.279$, and such results are consistent for other values of $E_d,V$: $(E_d+\ell)>0$.
Thus $\mu$ is negative.
We reiterate that $\mu$ can be determined analytically if the filling is integer.

\section{\label{app:mu} Methods of solving the saddle point equations}

Here we comment on the self-consistency procedure used to obtain the parameters $\mu,~r,~\ell$ in Ref.~\onlinecite{WKSM_PNAS} and this work.
Since the WKSM system was separable into pseudospin sectors, the Bloch Hamiltonian matrix was decomposed from one $8\times8$ to two $4\times4$ matrices.
A matrix size of $4\times4$ yields a characteristic equation with an eigenvalue polynomial degree of four, which is the upper limit to an exactly solvable eigenvalue problem.
When the calculation is exactly at quarter filling, the chemical potential does not have to be numerically determined, as shown in Appendix~\ref{app:pmD}.
However, for the calculations with a finite Fermi surface (e.g., Fig.~\ref{fig:berry}), we can determine the chemical potential $\mu$ numerically.

The remaining two saddle point equations are solved using the Newton-Raphson method.~\cite{NRF90:Press:1996} 
This method can solve nonlinear systems of equations, given an initial guess that is close enough to the eventual solution.
Whenever $r,~\ell$ are changed, $\mu$ is updated.
If the filling is specified precisely at the nodes (integer), $\mu$ is defined in Eq.~(\ref{eq:mu}). 
If the filling is noninteger, \textit{i.e.} metallic, it may be solved for by using the bisection method on the density $n(\mu)$ of particles per site per spin.


%

\end{document}